\documentclass[]{pasj01}

\usepackage{graphicx}
\usepackage{ccaption}
\usepackage{lscape}
\usepackage{comment}
\usepackage{threeparttable}
\usepackage{caption}
\usepackage{tablefootnote}
\usepackage[whole]{bxcjkjatype}
\usepackage{lineno}
\usepackage{multirow}
\draft

\begin{document} 
%\Received{}%{yyyy/mm/dd}
%\Accepted{}%{yyyy/mm/dd}
%\Published{yyyy/mm/dd}

\title{Distance determination of molecular clouds in the 1st quadrant of the Galactic plane using deep learning : I. Method and Results}

\author{Shinji \textsc{FUJITA}\altaffilmark{1,2}}%
\author{Atsushi \textsc{M.} \textsc{ITO}\altaffilmark{3}}%
\author{Yusuke \textsc{MIYAMOTO}\altaffilmark{4}}%
\author{Yasutomo \textsc{KAWANISHI}\altaffilmark{5}}%
\author{Kazufumi \textsc{TORII}\altaffilmark{6}}%
\author{Yoshito \textsc{SHIMAJIRI}\altaffilmark{7}}%
\author{Atsushi \textsc{NISHIMURA}\altaffilmark{8}}%
\author{Kazuki \textsc{TOKUDA}\altaffilmark{9,6}}%
\author{Toshikazu \textsc{OHNISHI}\altaffilmark{2}}%
\author{Hiroyuki \textsc{KANEKO}\altaffilmark{10,6}}%
\author{Tsuyoshi \textsc{INOUE}\altaffilmark{11}}%
\author{Shunya \textsc{TAKEKAWA}\altaffilmark{12}}%
\author{Mikito \textsc{KOHNO}\altaffilmark{13, 14}}%
\author{Shota \textsc{UEDA}\altaffilmark{2}}%
\author{Shimpei \textsc{NISHIMOTO}\altaffilmark{2}}%
\author{Ryuki \textsc{YONEDA}\altaffilmark{2}}%
\author{Kaoru \textsc{NISHIKAWA}\altaffilmark{14}}%
\author{Daisuke \textsc{YOSHIDA}\altaffilmark{14}}%

\altaffiltext{1}{Institute of Astronomy, Graduate School of Science, The University of Tokyo, 2-21-1 Osawa, Mitaka, Tokyo 181-0015, Japan} 
\altaffiltext{2}{Department of Physical Science, Graduate School of Science, Osaka Prefecture University 1-1 Gakuen-cho, Naka-ku, Sakai, Osaka 599-8531, Japan}
\altaffiltext{3}{National Institute for Fusion Science (NIFS), National Institutes of Natural Sciences (NINS) 322-6, Oroshi-cho, Toki, Gifu 509-5292, Japan}
\altaffiltext{4}{Department of Electrical and Electronic Engineering, Faculty of Engineering, Fukui University of Technology, 3-6-1, Gakuen, Fukui, Fukui, 910-8505, Japan}
\altaffiltext{5}{RIKEN Information R\&D and Strategy Headquarters, 2-2-2 Hikaridai, Seika-cho, Soraku-gun, Kyoto 619-0288, Japan}
\altaffiltext{6}{National Astronomical Observatory of Japan, National Institutes of Natural Sciences, 2-21-1 Osawa, Mitaka, Tokyo 181-8588, Japan}
\altaffiltext{7}{Kyushu Kyoritsu University, Jiyugaoka 1-8, Yahatanishi-ku, Kitakyushu, Fukuoka, 807-8585 Japan}
\altaffiltext{8}{Nobeyama Radio Observatory, National Astronomical Observatory of Japan, National Institutes of Natural Sciences, 462-2 Nobeyama, Minamimaki, Minamisaku, Nagano 384-1305, Japan}
\altaffiltext{9}{Department of Earth and Planetary Sciences, Faculty of Sciences, Kyushu University, Nishi-ku, Fukuoka 819-0395, Japan}
\altaffiltext{10}{Graduate School of Education, Joetsu University of Education, 1, Yamayashiki-machi, Joetsu, Niigata 943-8512, Japan}
\altaffiltext{11}{Department of Physics, Konan University, 8-9-1 Okamoto, Higashinada-ku, Kobe, Hyogo 658-8501, Japan}
\altaffiltext{12}{Faculty of Engineering, Kanagawa University, 3-27-1 Rokkakubashi, Kanagawa-ku, Yokohama, Kanagawa 221-8686, Japan}
\altaffiltext{13}{Astronomy Section, Nagoya City Science Museum, 2-17-1 Sakae, Naka-ku, Nagoya, Aichi 460-0008, Japan}
\altaffiltext{14}{Department of Astrophysics, Nagoya University, Furo-cho, Chikusa-ku, Nagoya, Aichi 464-8602, Japan}

\email{fujita.shinji@ioa.s.u-tokyo.ac.jp}


\maketitle

\begin{abstract}
%Is machine learning a viable tool for determining the distance to the molecular clouds in a galaxy?
Machine learning has been successfully applied in varied field but whether it is a viable tool for determining the distance to molecular clouds in the Galaxy is an open question. 
In the Galaxy, the kinematic distance is commonly employed to represent the distance to a molecular cloud. 
However, for the inner Galaxy, two different solutions, i.e., the ``Near'' solution and the ``Far'' solution, can be derived simultaneously. 
We attempt to construct a two-class (``Near'' or ``Far'') inference model using a convolutional neural network (CNN), which is a form of deep learning that can capture spatial features generally. 
In this study, we use the CO dataset in the 1st quadrant of the Galactic plane obtained with the Nobeyama 45-m radio telescope ($l=62^{\circ}-10^{\circ}$, $|b|<1^{\circ}$).
In the model, we apply the three-dimensional distribution (position--position--velocity) of the $^{12}$CO ($J=1-0$) emissions as the main input.
To train the model, a dataset with ``Near'' or ``Far'' annotation was created from the H{\sc ii} region catalog of the infrared astronomy satellite WISE.
Consequently, we construct a CNN model with a 76\% accuracy rate on the training dataset. 
Using the proposed model, we determine the distance to the molecular clouds identified by the CLUMPFIND algorithm.
We found that the mass of molecular clouds with a distance of $<\,8.15$\,kpc identified in the $^{12}$CO data follows a power-law distribution with an index of approximately $-2.3$ in the mass range $M\,>\,10^3\,M_{\odot}$.
In addition, the detailed molecular gas distribution of the Galaxy, as seen from the Galactic North pole, was determined.
\end{abstract}
%\linenumbers

\section{Introduction}
Stars are formed in molecular clouds.
\textcolor{black}{To better understand the initiation of star formation, investigating the physical properties of molecular clouds at various scales, from several hundred pc to sub pc, is vital. In addition, it is critical to study them throughout a galaxy} \textcolor{black}{because star formation is closely related to the structure and environment of galaxies.}
Recent advances in telescope technology have enabled wide-area CO surveys of many nearby galaxies [e.g., COMING (CO Multiline Imaging of Nearby Galaxies \cite{2016PASJ...68...89M, 2019PASJ...71S..14S}) and PHANGS (Physics at High Angular resolution in Nearby GalaxieS \cite{2021MNRAS.502.1218R, 2021ApJS..257...43L})]. 
The Atacama Large Millimeter/submillimeter Array (ALMA) has enabled the observation of M33 and the Large Magellanic Cloud/Small Magellanic Cloud with a spatial resolution of less than 1\,pc, facilitating the study of the global structure of galaxies and the relationship between gas dynamics and star formation (e.g., \cite{2020ApJ...896...36T, 2019ApJ...886...14F}).
\textcolor{black}{The spatial resolution of $<\,1\,$pc revealed filamentary structures of molecular clouds, the origin of which is currently under active discussion.}
The advances in telescope technology have also facilitated CO surveys of the Galaxy with higher angular resolution ($\sim \, 20''$ typically) and wide area ($\sim \, 100$ \\,degree$^2$ typically) [e.g., COHRS (CO High-Resolution Survey; \cite{2013ApJS..209....8D}), FUGIN (FOREST Unbiased Galactic plane Imaging survey with the Nobeyama 45-m telescope; \cite{2017PASJ...69...78U})\footnote{https://nro-fugin.github.io/}, SEDIGISM (Structure, Excitation, and Dynamics of the Inner Galactic Interstellar Medium; \cite{2021MNRAS.500.3064S})].
The FUGIN observations toward the Galactic plane revealed the presence of many cloud--cloud collisions triggering high-mass star formation (e.g., \cite{2020MNRAS.496.1278D, 2019ApJ...872...49F, 2021PASJ...73S.172F, 2018PASJ...70S..50K, 2018PASJ...70S..42N, 2018PASJ...70S..51T}).
Furthermore, the dense gas mass fraction in the Galactic plane was measured with high angular resolution (\cite{2019PASJ...71S...2T}), although the analysis was limited to nearby tangential regions because of the uncertainty of the distance to molecular clouds.

\textcolor{black}{For several decades, three-dimensional (3D) maps (or face-on maps) of the Galaxy have been a classic and fundamental topic.
The most reliable data for distances and 3D maps of the Galaxy are observations of masers using very long baseline interferometry (VLBI) (e.g., \cite{2014ApJ...783..130R}).
These observations provide strong evidence of the existence of spiral arms in the Galaxy. 
However, the number of data points is limited, and determining the extent to which the molecular clouds are associated is difficult.}
The distance information of molecular clouds in the Galaxy is a significant parameter not only because of its mass and size, but also because it enables discussion of the Galaxy's structure. 
The kinematic distance computed from the gas's line-of-sight velocity and the Galaxy's rotating velocity has been commonly employed; however, two different solutions, the ``Near'' solution and the ``Far'' solution, can be derived simultaneously for the gas in the inner solar system orbit (this is termed as the Near--Far problem). 
Several methods have been proposed to solve this problem and to determine the distance (e.g., \cite{2006PASJ...58..847N, 2020A&A...640A..72R, 2021A&A...646A..74M}).
For example, assuming that the vertical distribution of H$_2$ gas follows the equation \textcolor{black}{(${\rm sech}^2$ function obtained by treating it as an isothermal and self-gravitating system)} in \citet{1942ApJ....95..329S}, \citet{2006PASJ...58..847N} divided the CO data obtained with a CfA 1.2-m Millimeter-Wave Telescope (\cite{2001ApJ...547..792D}) into ``Near'' and ``Far'' emissions; using this, they presented a 3D map of the molecular gas of the Galaxy. 
\citet{2020A&A...640A..72R} adopted a Bayesian approach (\cite{2016ApJ...823...77R, 2019ApJ...885..131R}) to derive the current best assessment of the Galactic distribution of $^{13}$CO from the Galactic Ring Survey (GRS) (\cite{2006ApJS..163..145J}).

Machine-learning techniques, particularly deep learning, are widely accepted as powerful tools in various fields.
Deep learning has been particularly successful in the field of imaging, such as detection of disease in the medical field and typhoon in meteorology (e.g., \cite{Sumaiya, Matsuoka}).
In astronomy, many studies have used machine learning, such as those on the morphological classification of galaxies and anomaly detection of signals (e.g., \cite{2021MNRAS.507.1937B, 2021ApJS..255...24V}).
\citet{2020SPIE11452E..2LU} attempted to identify infrared rings, which have been identified only by the human eye, using an object detection model based on a convolutional neural network (CNN; e.g., \cite{2014arXiv1404.7828S}), which is a form of deep learning.
They succeeded in developing a model that was comparable to human eyes.

In this paper (Paper I), our main motivation is to label all voxels of the CO data cube (position--position--velocity) as ``Near'' or ``Far'' using a CNN model. 
The kinematic distance to molecular clouds is then determined by combining the labels with the rotation parameter of the Galaxy; the physical properties of molecular clouds in the 1st quadrant of the Galactic plane can be observed with unprecedented high spatial resolution.
In Section\,2, we present the CO data, whereas in Section\,3, we describe the CNN model.
Next, in Section\,4, we present cloud identification and distance determination.  
\textcolor{black}{In Section\,5, we show the $^{13}$CO face-on-view map of the Galaxy, whereas in Section\,6, we highlight possible errors in distance estimation in this study.}
In the forthcoming paper (Paper II), we will discuss the physical properties of the molecular clouds, such as the dense gas mass fraction and the Galactic structure. 
In this study, we assume that the rotation curve of the Galaxy is flat; moreover, we assume that the distance from the Sun to the Galactic center is $8.15$\,kpc, and adopt a rotation speed of $236$\,km\,s$^{-1}$ (\cite{2019ApJ...885..131R}).

\section{Data}
We used FUGIN $^{12}$CO, $^{13}$CO, and C$^{18}$O ($J=1-0$) emission data obtained using the Nobeyama 45-m radio telescope ($l=50^{\circ}-10^{\circ}$, $|b|< 1^{\circ}$; \cite{2017PASJ...69...78U}). 
See \citet{2017PASJ...69...78U} for details of the observations.
We downloaded version 1.00, which fits the data cube from the archive site\footnote{http://jvo.nao.ac.jp/portal/nobeyama/fugin.do}.
We also used the $^{12}$CO, $^{13}$CO, and C$^{18}$O ($J=1-0$) emission data obtained with the Nobeyama 45-m radio telescope ($l=62^{\circ}-50^{\circ}$, $|b|< 1^{\circ}$; \cite{2022PASJ...74...24K}; Nishimura et al. in prep.).
These observations were conducted using a scan mode that is similar to FUGIN, and their effective spatial resolutions are $\sim \,20''$. 
The velocity coverage ranges from $-100$ to $+200$\,km\,s$^{-1}$.
In this study, the data cubes were spatially convoluted with $30''$ Gaussian (effective angular resolution of $\sim 36''.1$) to enhance the signal-to-noise ratio (S/N) and to remove the scanning effect. 
Figure\,\ref{fig:lb} shows the peak brightness temperature map and the $T_{\rm rms}$ map of the three lines. 
The typical \textcolor{black}{noise level} ($T_{\rm rms}$) is $1.0-1.5$\,K, $0.5-0.8$\,K, and $0.5-0.8$\,K for $^{12}$CO, $^{13}$CO, and C$^{18}$O ($J=1-0$) emissions in \textcolor{black}{the main beam brightness temperature} ($T_{\rm mb}$) scale, respectively, although the $T_{\rm rms}$ map in Figure\,\ref{fig:lb} is not uniform. 
The angular and velocity grid sizes were $8''.5$ and $0.65$ km\,s$^{-1}$, respectively. 
The FITS cubes were created with a size of $2^{\circ}\, \times \,2^{\circ}\, \times \,300$\,km\,s$^{-1}$ ($848$ \, pixels\,$\times$\,$848$\,pixels\,$\times$\,$462$\,channels) every $1^{\circ}$ in the Galactic longitude (51\, FITS cubes for each emission line).

\begin{landscape}
\begin{figure}
 \begin{center}
  \includegraphics[width=\hsize]{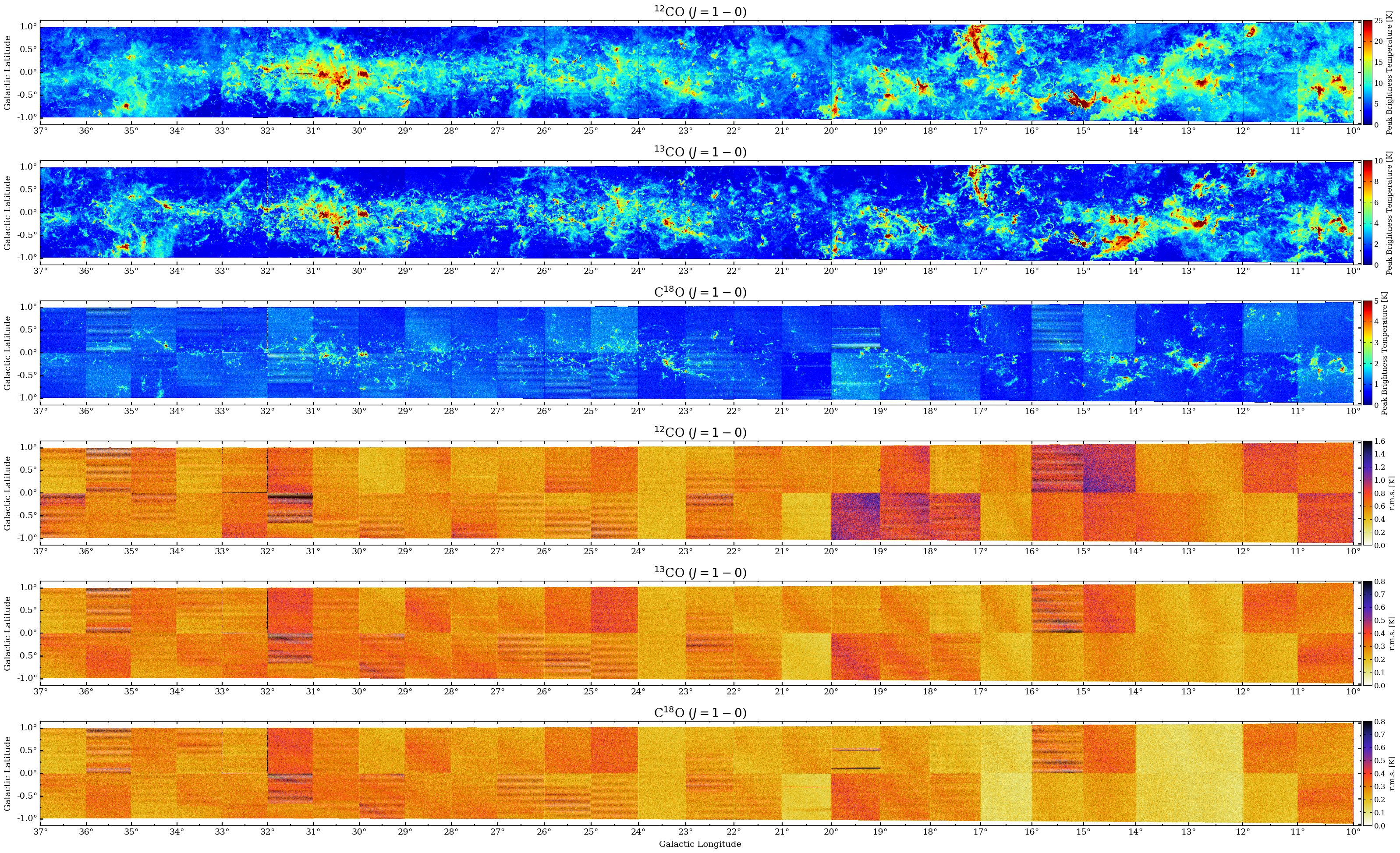} 
 \end{center}
\caption{Peak brightness temperature map and $T_{rms}$ map of the $^{12}$CO, $^{13}$CO, and C$^{18}$O ($J=1-0$) emission.}\label{fig:lb}
\end{figure}
\end{landscape}

\begin{landscape}
\begin{figure}
 \begin{center}
  \includegraphics[width=\hsize]{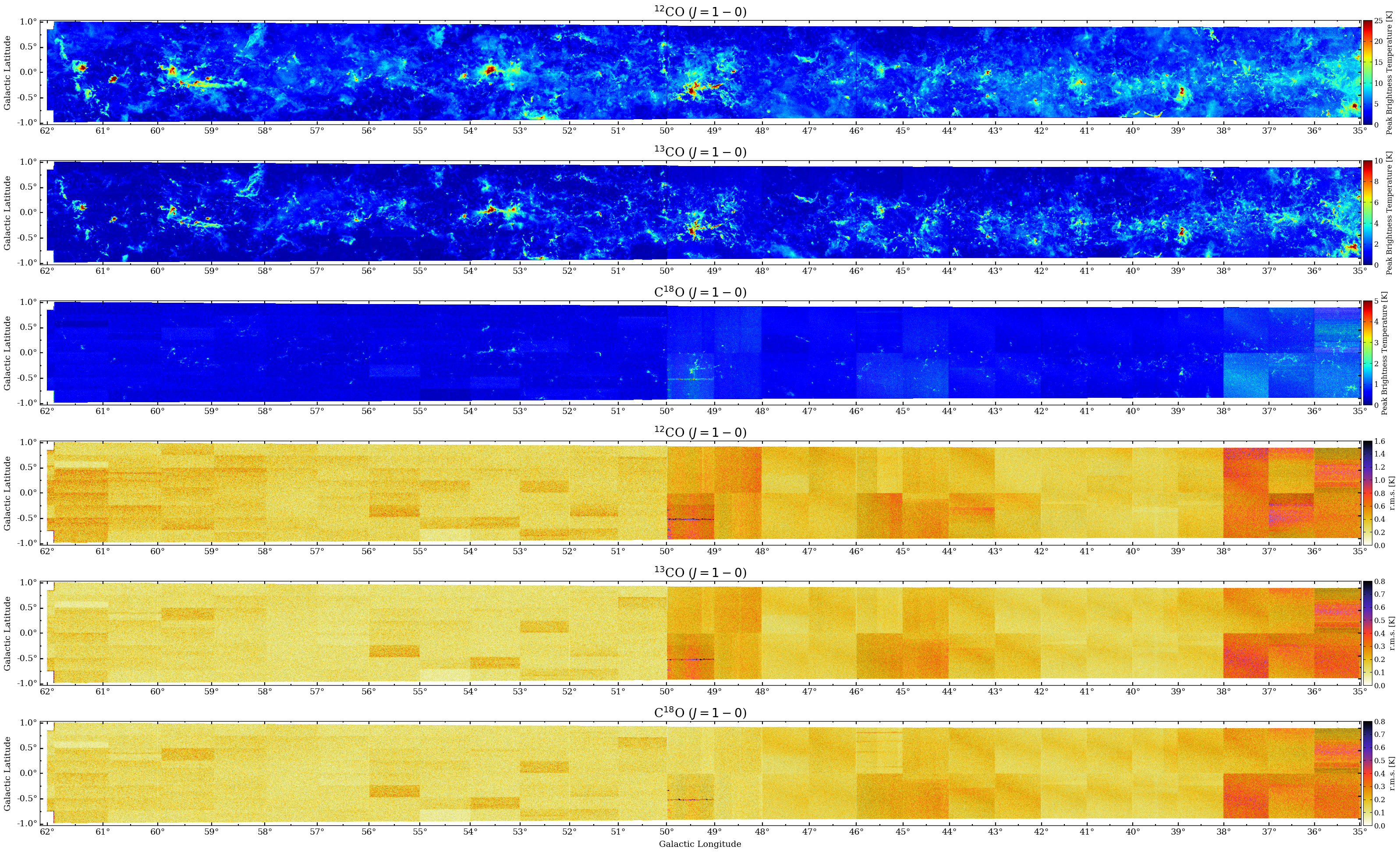} 
 \end{center}
\contcaption{(Continued)}
\end{figure}
\end{landscape}

\section{Near-Far Labeling with a Convolutional Neural Network Model}\label{sec:model}

\subsection{Model architecture}
\begin{figure}
 \begin{center}
  \includegraphics[width=\hsize]{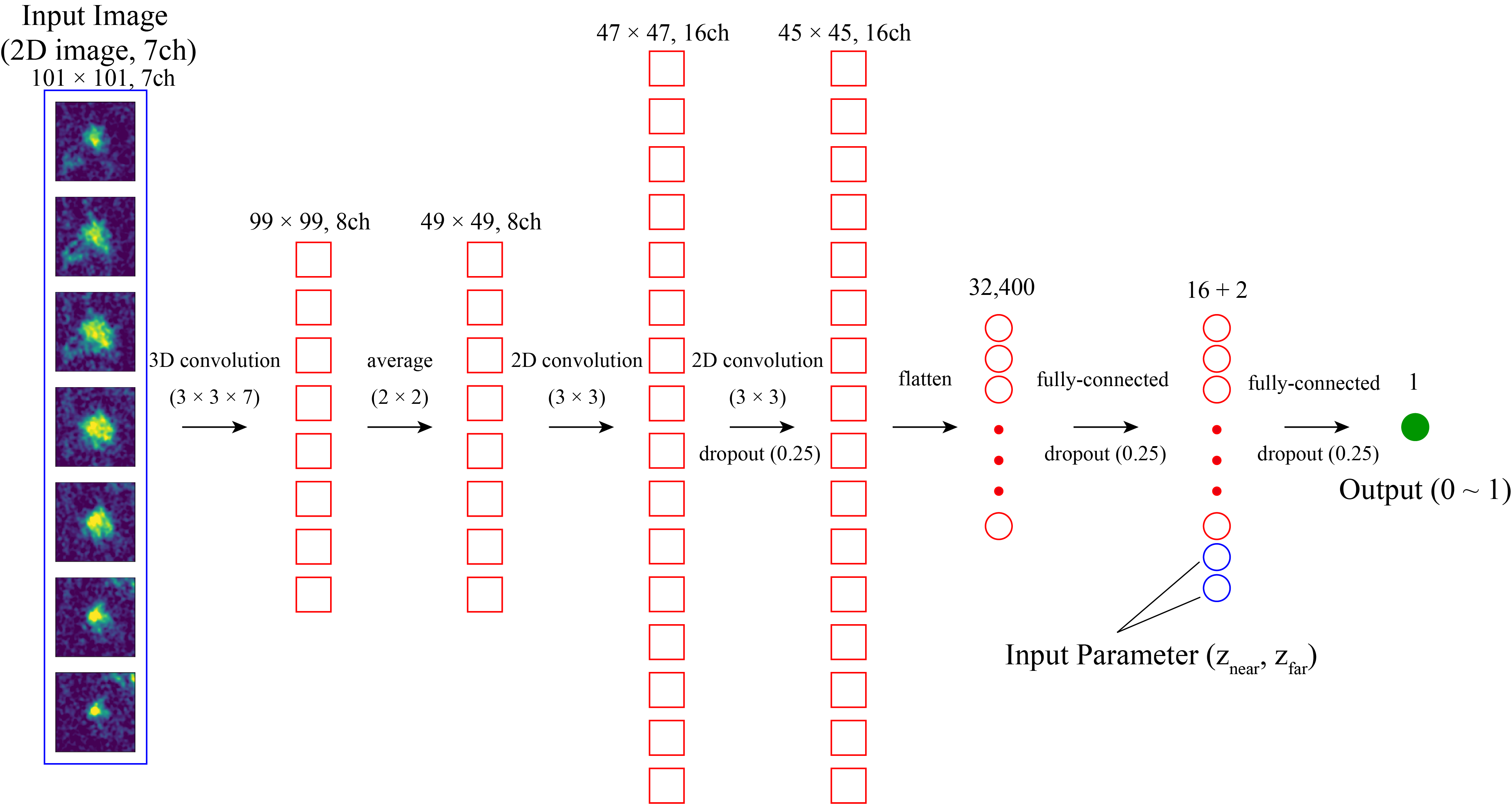} 
 \end{center}
\caption{ CNN's architecture. Blue, red, and green indicate input data, feature values, and output, respectively. }\label{fig:CNN}
\end{figure}

We propose a Near-Far decision model based on a CNN, which is generally strong in image recognition. 
% Tensorflow, a Python package, was employed to develop the model.
The model's input data consist of two types: sensory data $\mathbf{X}_i$ and a two-dimensional (2D) vector $\mathbf{z}_i$.
The sensory data $\mathbf{X}_i$ indicate a ${}^{12}$CO ($J=1-0$) cube ($101 \mbox{ spatial pixels} \times 101 \mbox{ spatial pixels} \times 7 \mbox{ velocity channels}$).
The 2D vector $\mathbf{z}_i$ consists of two values, $z_{\rm near}$ and $z_{\rm far}$, which indicate the floats from the Galactic disk ($D\,\sin\,b$, where $D$\,[kpc] and $b$\,[$^{\circ}$] are the distance and the Galactic latitude, respectively) when the inputted cloud is ``Near'' and ``Far,'' respectively.
Using $\mathbf{z}_i$ as an input, we were able to suppress the presence of clouds that were too far from the Galactic disk, thus slightly enhancing the accuracy of the model for the training dataset.
The proposed CNN model can be formulated as follows.
\begin{equation}
y_i = f(\mathbf{X}_i, \mathbf{z}_i; \mathbf{\Theta}),
\end{equation}
where $\mathbf{\Theta}$ denotes the set of parameters to be trained.
Figure\,\ref{fig:CNN} shows in detail the architecture of the CNN model $f$.
The CNN model $f$ consists of three convolution layers, followed by two fully connected layers.
We inserted an average pooling layer after the first convolution layer to reduce the input size and make the model more robust.
We added ``dropout'' to suppress overfitting. 
The ReLU function and the Sigmoid function were used as activation functions for the middle and final fully connected layers, respectively.
The output is a single value, $y_i\in[0, 1]$, which indicates whether the input is ``Near'' or ``Far.''
This model was implemented using the Python package Tensorflow, and the total number of parameters $\mathbf{\Theta}$ was 522,435.
We minimized this loss when training the CNN. 
\begin{equation}
L(y_i, \widehat{y}_i) = - \widehat{y}_i \log y_i,
\end{equation}
where $\widehat{y}_i\in\{0, 1\}$ represents the binary-annotated values of the training data.
The parameter set $\mathbf{\Theta}$ was optimized through training.

%Figure\,\ref{fig:CNN} shows the architecture of the model. 
%The input data is a $^{12}$CO ($J=1-0$) cube (101 spatial pixels $\times$ 101 spatial pixels $\times$ 7 velocity channels) and two scalars, $z_{near}$ and $z_{far}$, which are floats from a galaxy disk when the inputted cloud is Near and Far, respectively. 
%These input data are converted as a single decimal number (from $0$ to $1$), which corresponds to ``Near'' (0) or ``Far'' (1), through the layers in the model. 
%We added ``dropout'' to suppress overfitting and to activate all perceptrons.
%The ReLU function and the Sigmoid function were used as activation functions for the middle layers and the final dense layer, respectively.
%%Binary cross-entropy was used as the loss function.
%The total number of parameters was 522,435. 

\subsection{Molecular clouds for the dataset with Near--Far annotations}
The molecular clouds listed in the WISE H{\sc ii} region catalog\footnote{http://astro.phys.wvu.edu/wise/} (\cite{2014ApJS..212....1A}) as the dataset. 
Among the H{\sc ii} regions located in region ($l=61^{\circ}.8-10^{\circ}.2$, $|b|<0.^{\circ}8$), we selected only those that satisfied the following conditions: 
1) H{\sc ii} regions with a label of ``Near'' or ``Far,'' 
2) H{\sc ii} regions with associated $^{12}$CO ($J=1-0$) emissions; 
3) H{\sc ii} regions with a ratio of far distance to near distance greater than two, i.e., regions other than those near the tangent.
In addition, several local clouds with line-of-sight velocities of $0-10$\,km\,s$^{-1}$ were added as ``Near'' clouds.
The number of ``Near'' and ``Far'' molecular clouds was 91 and 159, respectively.

\subsection{Training of model}
\begin{figure}
 \begin{center}
  \includegraphics[width=15cm]{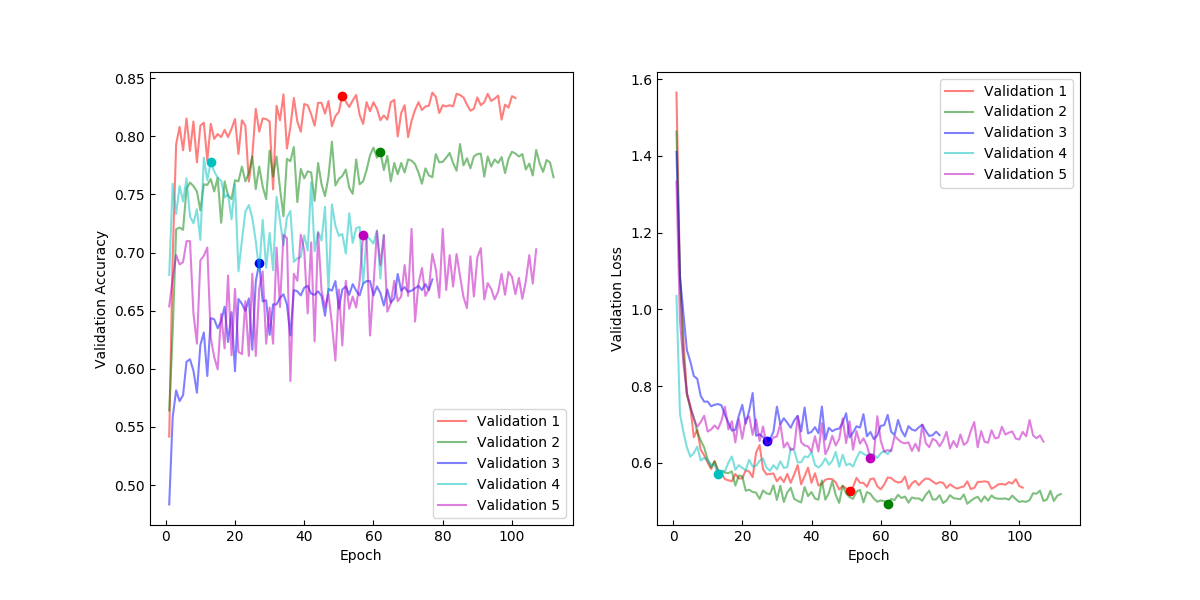} 
 \end{center}
\caption{Variation of the models' ``Accuracy'' and ``Loss'' (cross-validation). The five dots indicate the saved point when the validation loss is the minimum. }\label{fig:vals}
\end{figure}

We randomly divided the dataset into five groups (four training datasets and one validation dataset), after which we trained and evaluated the model using cross-validation (Figure\,\ref{fig:vals}). 
After dividing the original dataset, the training sets were augmented (random rotation, random flipping, and addition of a random small offset) up to a total of 10,000 samples.

At every epoch, the sum of the loss values over the samples in the validation set (validation loss) was monitored.
If no improvement was made for more than 50 epochs, the training was terminated, and the model parameters at the checkpoint (dot markers in Figure\,\ref{fig:vals}) with the minimum validation loss were saved.
The average percentage of correct answers in the validation data throughout the training was approximately 76\%. 
%This value is about 10\% higher than when using other machine learning models. Separately, Random Forest (trained the physical parameters of molecular clouds such as intensity of emissions, size, and line width)  was performed.

\subsection{Distance determination for each voxel}\label{sec:Distance}
\begin{figure}
 \begin{center}
  \includegraphics[width=15cm]{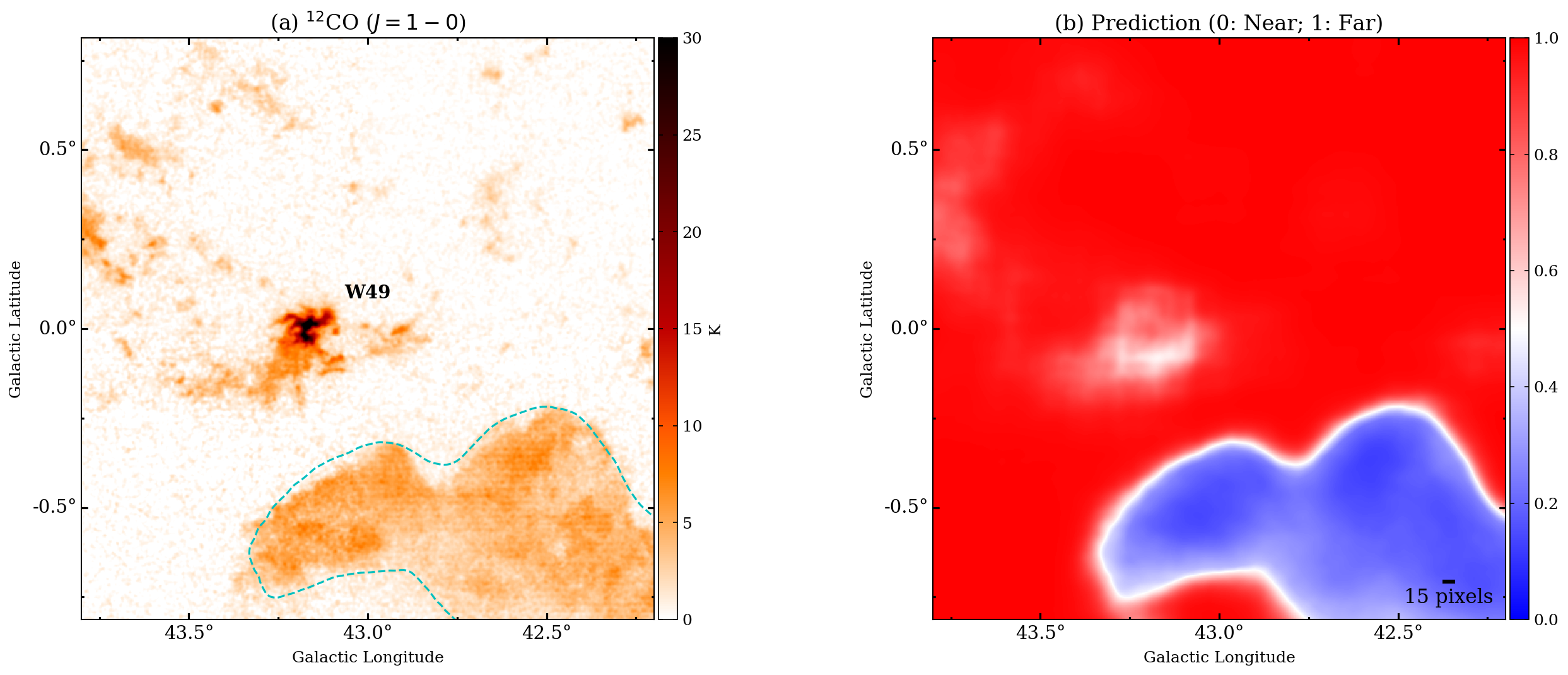} 
 \end{center}
\caption{(a) Brightness temperature of the $^{12}$CO ($J=1-0$) emission, ($v_{\rm LSR}=9.5$\,km\,s$^{-1}$) toward W49. The cyan dotted-line indicates where the predicted value is 0.5 in (b). (b) Predicted value of Near--Far by the CNN model. The values 0 and 1 refer to ``Near'' and ``Far,'' respectively. }\label{fig:test_reg}
\end{figure}

We applied the five models to all voxels that required inference, that is, $v_{\rm LSR}\,>\,0$\,km\,s$^{-1}$ (inside the orbit of the solar system for the 1st quadrant). 
To reduce the computational cost, the inference was performed every seven pixels in the spatial direction (\textcolor{black}{the computational cost was reduced to $\sim \,1/50$}).
Seven pix is sufficiently small for the input data shape (101 $\times$ 101), and \textcolor{black}{by comparing the 7-pix sampled result with the full sampled result shown in Figure\,\ref{fig:test_reg}(b)}, we confirmed that it does not affect the inference accuracy.
We used the averaged inference value of the five models. 

Figure\,\ref{fig:test_reg}(a) shows the brightness temperature of the $^{12}$CO ($J=1-0$) emission ($v_{\rm LSR}=9.5$\,km\,s$^{-1}$) toward W49.
There are $^{12}$CO emissions from both the W49 molecular clouds and a nearby molecular cloud in this velocity channel.
Figure\,\ref{fig:test_reg}(b) shows the predicted value of Near--Far using the CNN model. 
0 and 1 refer to ``Near'' and ``Far,'' respectively.
%In this paper, voxels with values less than or equal to $0.5$ are considered as ``Near'' and voxels with values greater than $0.5$ are considered as ``Far.''
In this study, by binarizing the output value of the trained model using a threshold of $0.5$, we can obtain the result of the Near-Far decision for the input. 
Specifically, if the output value is lower than the threshold of $0.5$, the final decision is ``Near''; otherwise, the final decision is ``Far.''
The area considered to be ``Near'' is the area inside the dotted line cyan contour in Figure\ref{fig:test_reg}(a).
For these figures, it was confirmed that the Near--Far separation was accurate at least in this region, although the edge of the local molecular cloud is determined to be ``Far.''
As shown in Section\,\ref{sec:dd}, this effect is largely eliminated by taking a majority vote on the voxels that compose the cloud for each.

%\clearpage

\section{Cloud Identification and Distance Determination}

\subsection{Cloud identification}\label{sec:ci}
Cloud identification was performed using CLUMPFIND in PyCupid (\cite{1994ApJ...428..693W, 2007ASPC..376..425B}) for $^{12}$CO, $^{13}$CO, and C$^{18}$O ($J=1-0$) cubes.
The identification parameter ``MINPIX'' (minimum number of voxels) was set to 500, 500, and 125 for $^{12}$CO, $^{13}$CO, and C$^{18}$O ($J=1-0$), respectively, and ``LEVEL'' (contour levels) were set to \textcolor{black}{[1.5\,K (typically $\sim \,1\,\sigma$), 5.0\,K ($\sim \,3\,\sigma$), 10.0\,K, 15.0\,K, ...], [1.0\,K (typically $\sim \,1\,\sigma$), 2.5\,K  ($\sim \,3\,\sigma$), 5.0\,K, 7.5\,K, ...], and [1.0\,K (typically $\sim \,1\,\sigma$), 2.0\,K (typically $\sim \,2\,\sigma$), 3.0\,K, 4.0\,K, ...]} for $^{12}$CO, $^{13}$CO, and C$^{18}$O ($J=1-0$), respectively.
The typical $T_{\rm rms}$ values of the three lines in the noisy regions are approximately 1.5\,K, 1.0\,K, and 1.0\,K for $^{12}$CO, $^{13}$CO, and C$^{18}$O ($J=1-0$), respectively.
Therefore, we set ``MINPIX'' to be higher than in typical use cases to prevent the detection of the noise as the cloud. 
\textcolor{black}{Note that the choice of these parameters may affect the results (the mass function in particular).}
\textcolor{black}{\citet{2009ApJ...699L.134P} proposed that the CLUMPFIND parameter ``stepsize'' has a significant impact on the mass function when applied to three-dimensional (3D) data.
In this study, a uniform, albeit arbitrary, parameter was applied although the noise levels and distances varied from region to region.}

For the C$^{18}$O data, the four regions $(l ({\rm degree}), \,b({\rm degree})) = (49.98$-$-49.96, \,-0.38$-$-0.35), (50.00$-$-49.00, \,-0.58$-$-0.54), (32.02$-$-32.00, \,0.00$-$+1.00), (33.01$-$-33.00, \,0.00$-$+1.00)$ were removed because the noise level was too high ($>\,5$\,K typically).
As a result, 142933, 37963, and 6664 clouds were identified in the $^{12}$CO, $^{13}$CO, and C$^{18}$O ($J=1-0$) data cubes, respectively (hereafter referred to as $^{12}$CO clouds, $^{13}$CO clouds, and C$^{18}$O clouds).
The identified clouds are listed in Tables\,1, 2, and 3.

\subsection{Distance determination for each identified cloud}\label{sec:dd}
\begin{landscape}
\begin{figure}
 \begin{center}
  \includegraphics[width=\hsize]{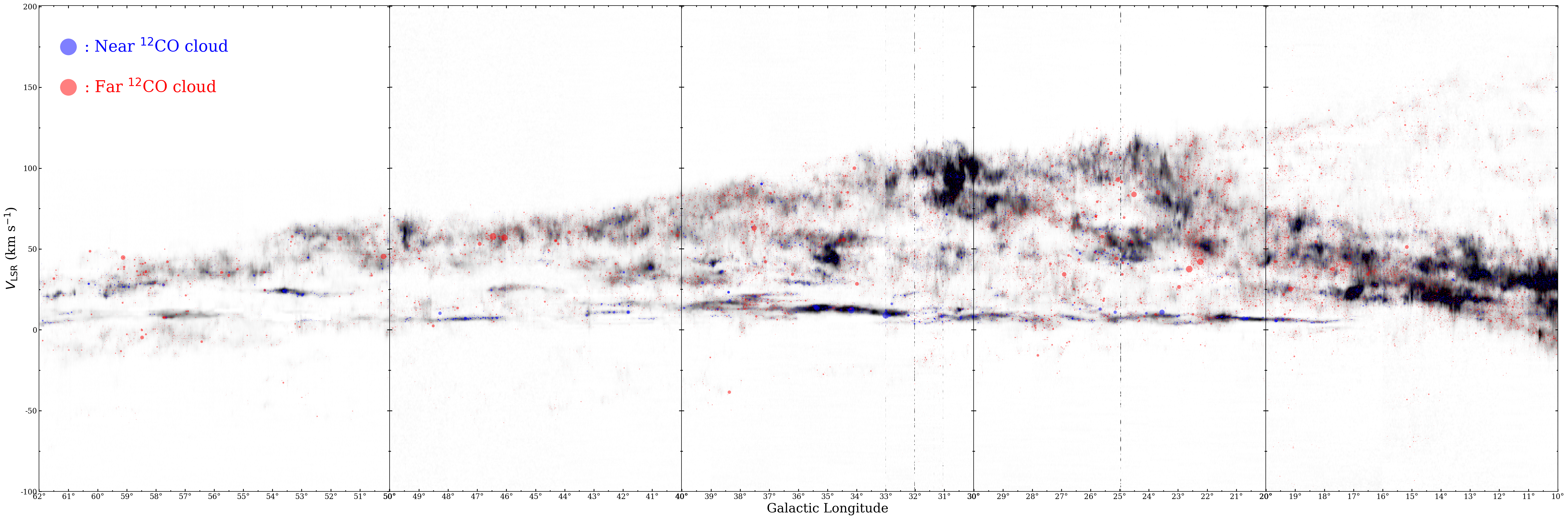} 
 \end{center}
\caption{$l$-$v$ diagram of the $^{12}$CO ($J=1-0$) emission. The circles represent the position of the $^{12}$CO clouds. The size of the circle is proportional to the area of the cloud. The blue and red circles refer to the ``Near'' and ``Far'' clouds, respectively. }\label{fig:lvsum_12CO_poster_5}
\end{figure}
\end{landscape}

\begin{figure}
 \begin{center}
  \includegraphics[width=15cm]{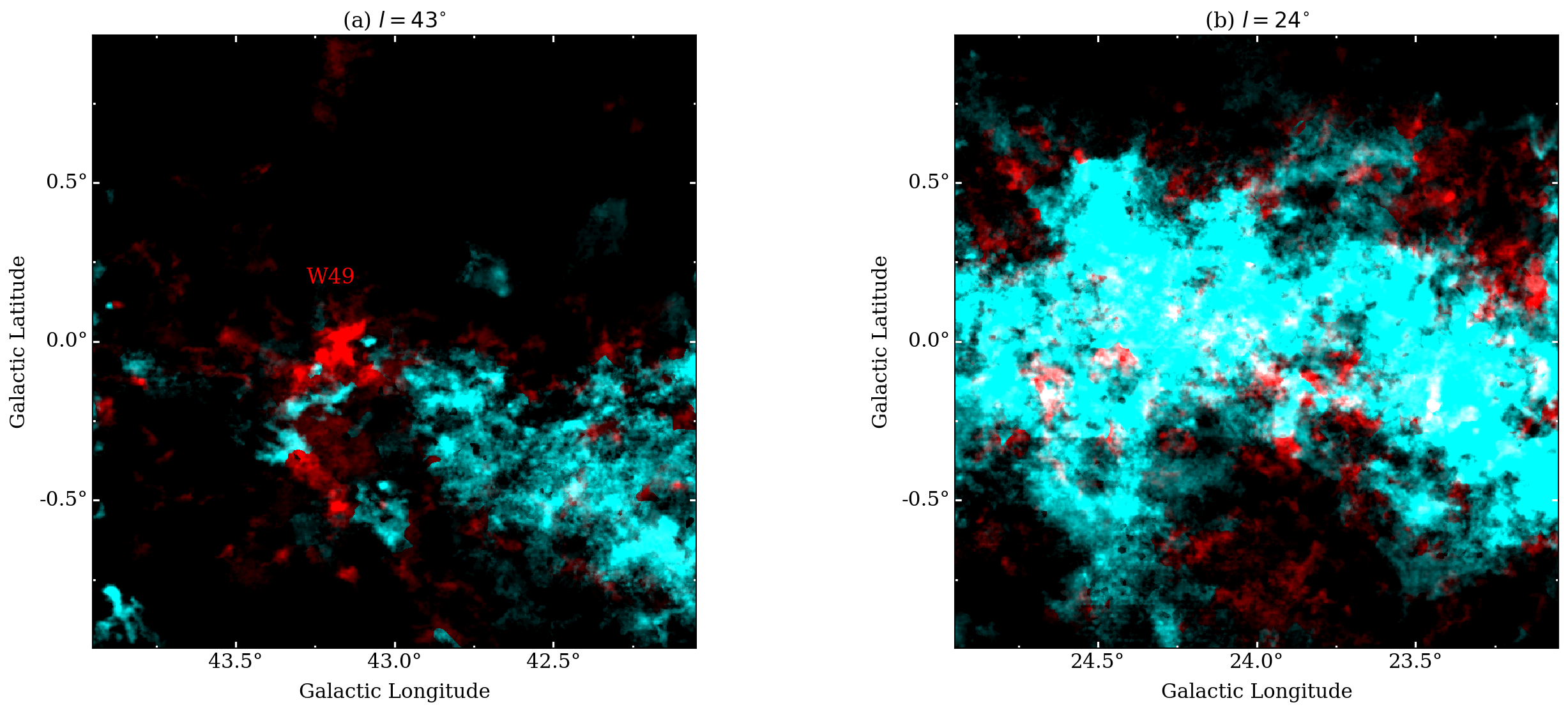} 
 \end{center}
\caption{\textcolor{black}{Integrated intensity map of the $^{13}$CO ($J=1-0$) emission toward the (a) $l=43^{\circ}$ region and (b) $l=24^{\circ}$ region. The cyan and red clouds indicate the ``Near'' clouds and ``Far'' clouds, respectively.}}\label{fig:RB_map}
\end{figure}

Distance information is required to determine the mass of the identified cloud.
For each identified cloud, we listed the CNN inference values (Section\,\ref{sec:Distance}) of the voxels contained in the cloud and counted the number of values greater than 0.5.
If the counted number was a majority of the total number of voxels, we consider the cloud as ``Far;'' otherwise, it was considered ``Near.''
Kinematic distances were assigned to all identified clouds using this procedure.
Figure\,\ref{fig:lvsum_12CO_poster_5} shows the $l$–$v$ diagram of the $^{12}$CO ($J=1-0$) emission. 
The circles represent the position of the $^{12}$CO clouds, whereas the blue and red represent the ``Near'' and ``Far'' cloud, respectively.
\textcolor{black}{Figure\,\ref{fig:RB_map} shows the integrated intensity map of the $^{13}$CO ($J=1-0$) emission toward sample regions, (a) the $l=43^{\circ}$ region and (b) the $l=24^{\circ}$ region. 
The cyan and red clouds indicate the ``Near'' and ``Far'' clouds, respectively. 
The results of a more detailed analysis of these molecular clouds are discussed in Paper II.}

\clearpage
\begin{landscape}
\begin{center}
\begin{table}
  \tbl{Molecular clouds identified in the $^{12}$CO ($J=1-0$) data}{
  \label{tab:12CO}
  \begin{tabular}{c|cccccccccccccccc}
  \hline              
  ID & $l$ & $b$ & $v_{\rm LSR}$ & $n_{\rm vox}$	 & $\sigma _v$ & $T_{\rm ^{12}CO}$ & $T_{\rm ^{13}CO}$ & $T_{\rm C^{18}O}$ & $w_{\rm ^{12}CO}$ & $w_{\rm ^{13}CO}$ & $w_{\rm C^{18}O}$ & NF & $D$ & $M_{\rm LTE}$ & $R_{\rm gal}$ & $s$\\
  & [$^{\circ}$] & [$^{\circ}$] & [km\,s$^{-1}$] &  & [km\,s$^{-1}$] & [K] & [K] & [K] & [K\,km\,s$^{-1}$ & [K\,km\,s$^{-1}$ & [K\,km\,s$^{-1}$ &  & [kpc] & [M$_{\odot}$] & [kpc] & [pc]\\
  &   &   &   & &   &   &   &   &   & degree$^2$] & degree$^2$] & degree$^2$] &  &  &  &  \\
  \hline
  \hline
  12CO-000001 & 10.621 & -0.381 & -4.12 & 45358 & 3.00 & 47.6 & 24.0 & 5.0 & 1.69e+00 & 4.21e-01 & 7.42e-02 & F(1.00) & 16.89 & 1.53e+06 & 9.00 & 10.31 \\
12CO-000002 & 10.628 & -0.336 & -3.48 & 31397 & 2.97 & 42.3 & 21.7 & 8.0 & 1.14e+00 & 2.72e-01 & 4.86e-02 & F(1.00) & 16.74 & 8.94e+05 & 8.86 & 8.27 \\
12CO-000003 & 10.595 & -0.367 & -2.17 & 28521 & 2.86 & 41.7 & 15.2 & 4.5 & 1.02e+00 & 2.39e-01 & 3.54e-02 & F(1.00) & 16.46 & 7.33e+05 & 8.58 & 8.94 \\
  ... &   &   &   & &   &   &   &   &   &   &   &   &  &  &  &  \\
  12CO-141540 & 61.712 & 0.752 & 5.04 & 2633 & 0.54 & 5.0 & 0.6 & -0.1 & 2.14e-02 & 1.23e-03 & 2.78e-04 & N(0.37) & 0.43 & 8.60e-01 & 7.96 & 0.33 \\
12CO-141541 & 61.216 & 0.303 & 34.31 & 11376 & 1.35 & 5.0 & 0.4 & -0.2 & 9.28e-02 & 1.35e-02 & 2.34e-04 & F(0.97) & 3.92 & 8.29e+02 & 6.99 & 3.93 \\
12CO-141542 & 60.168 & -0.244 & 25.86 & 1212 & 1.16 & 5.0 & 0.4 & 0.3 & 1.12e-02 & 8.84e-04 & 6.23e-05 & F(0.95) & 5.60 & 1.09e+02 & 7.24 & 1.02 \\
  \hline
  \hline
\end{tabular}}
\end{table}
\begin{tablenotes}
\item[*] $l$, $b$, and $v_{\rm LSR}$ are the peak positions of the $^{12}$CO ($J=1-0$) emission. 
$n_{\rm vox}$ is the number of voxels that the cloud contains. 
$\sigma _v$ is the intensity-weighted standard deviation of the velocity. 
$T_{\rm ^{12}CO}$, $T_{\rm ^{13}CO}$, and $T_{\rm C^{18}O}$ are the respective intensities of the $^{12}$CO ($J=1-0$), $^{13}$CO ($J=1-0$), and C$^{18}$O ($J=1-0$) emission at the peak position.
$w_{\rm ^{12}CO}$, $w_{\rm ^{13}CO}$, and $w_{\rm C^{18}O}$ are the combined intensities of the three emissions, respectively.
NF refers to Near (N) or Far (F) assigned by us. 
\textcolor{black}{Decimals in parentheses indicate the confidence of the inference (median value of the inference by five models in all voxels); closer to 0.0 means ``Near'', and closer to 1.0 means ``Far''.}
$D$ is the kinematic distance to the cloud.
$M_{\rm LTE}$ is the mass of the cloud derived from LTE analysis using the peak brightness temperature of the $^{12}$CO ($J=1-0$) and the brightness temperature of the $^{13}$CO ($J=1-0$). We adopted an abundance ratio [$^{13}$CO]/[H$_2$] of $1.5 \times 10^{-6}$. 
$R_{\rm gal}$ is the Galactocentric distance to the cloud. 
\textcolor{black}{$s$ is the size of the cloud ($D\,\tan{(\sqrt{\sigma _l \sigma _b}})$, where $\sigma _l$ and $\sigma _b$ are the intensity-weighted standard deviation values along the Galactic Longitude axis and Galactic Latitude axis, respectively).}
(This table is available in its entirety in machine-readable form.)
\end{tablenotes}
\end{center}
\end{landscape}

\clearpage
\begin{landscape}
\begin{center}
\begin{table}
  \tbl{Molecular clouds identified in the $^{13}$CO ($J=1-0$) data.}{
  \label{tab:13CO}
  \begin{tabular}{c|cccccccccccccccc}
  \hline              
  ID & $l$ & $b$ & $v_{\rm LSR}$ & $n_{\rm vox}$	 & $\sigma _v$ & $T_{\rm ^{12}CO}$ & $T_{\rm ^{13}CO}$ & $T_{\rm C^{18}O}$ & $w_{\rm ^{12}CO}$ & $w_{\rm ^{13}CO}$ & $w_{\rm C^{18}O}$ & NF & $D$ & $M_{\rm LTE}$ & $R_{\rm gal}$ & $s$\\
  & [$^{\circ}$] & [$^{\circ}$] & [km\,s$^{-1}$] &  & [km\,s$^{-1}$] & [K] & [K] & [K] & [K\,km\,s$^{-1}$ & [K\,km\,s$^{-1}$ & [K\,km\,s$^{-1}$ &  & [kpc] & [M$_{\odot}$] & [kpc] & [pc] \\
  &   &   &   & &   &   &   &   &   & degree$^2$] & degree$^2$] & degree$^2$] &  &  &  &  \\
  \hline
  \hline
  13CO-000001 & 10.621 & -0.379 & -2.83 & 12370 & 2.09 & 46.2 & 26.6 & 7.9 & 8.41e-01 & 2.35e-01 & 4.07e-02 & F(1.00) & 16.60 & 8.44e+05 & 8.72 & 6.36 \\
13CO-000002 & 10.628 & -0.336 & -4.12 & 7263 & 1.73 & 41.9 & 22.9 & 8.1 & 5.18e-01 & 1.57e-01 & 3.26e-02 & F(1.00) & 16.89 & 5.40e+05 & 9.00 & 4.79 \\
13CO-000003 & 10.595 & -0.365 & -2.83 & 16032 & 1.85 & 40.2 & 18.0 & 4.3 & 8.39e-01 & 2.25e-01 & 3.32e-02 & F(1.00) & 16.60 & 7.09e+05 & 8.72 & 8.34 \\
  ... &   &   &   & &   &   &   &   &   &   &   &   &  &  &  &  \\
  13CO-037607 & 60.335 & -0.769 & 28.46 & 835 & 0.49 & 6.8 & 2.6 & 0.4 & 1.84e-02 & 4.59e-03 & 1.61e-04 & N(0.35) & 3.00 & 1.87e+02 & 7.16 & 0.62 \\
13CO-037608 & 61.726 & 0.575 & 20.65 & 683 & 0.55 & 9.2 & 2.5 & 0.3 & 1.68e-02 & 3.80e-03 & 1.99e-04 & N(0.25) & 2.01 & 7.32e+01 & 7.41 & 0.50 \\
13CO-037609 & 61.452 & -0.492 & 28.46 & 1625 & 0.89 & 5.2 & 2.5 & 0.2 & 2.58e-02 & 7.67e-03 & 3.47e-04 & F(0.95) & 4.21 & 5.92e+02 & 7.17 & 1.30 \\
  \hline
  \hline
\end{tabular}}
\end{table}
\begin{tablenotes}
\item[*] $l$, $b$, and $v_{\rm LSR}$ are the peak positions of the $^{13}$CO ($J=1-0$) emission.
$n_{\rm vox}$ is the number of voxels that the cloud contains. 
$\sigma _v$ is the intensity-weighted standard deviation of the velocity. 
$T_{\rm ^{12}CO}$, $T_{\rm ^{13}CO}$, and $T_{\rm C^{18}O}$ are the intensities of the $^{12}$CO ($J=1-0$), $^{13}$CO ($J=1-0$), and C$^{18}$O ($J=1-0$) emission, respectively, at the peak position.
$w_{\rm ^{12}CO}$, $w_{\rm ^{13}CO}$, and $w_{\rm C^{18}O}$ are respectively the combined intensities of the three emissions.
NF indicates Near (N) or Far (F), which we assigned. 
\textcolor{black}{Decimals in parentheses indicate the confidence of the inference (median value of the inference by five models in all of the voxels); closer to 0.0 means ``Near'', and closer to 1.0 means ``Far''.}
$D$ is the kinematic distance to the cloud. 
$M_{\rm LTE}$ is the mass of the cloud derived from LTE analysis using the peak brightness temperature of $^{12}$CO ($J=1-0$) and the brightness temperature of $^{13}$CO ($J=1-0$). We adopted an abundance ratio [$^{13}$CO]/[H$_2$] of $1.5 \times 10^{-6}$. 
$R_{\rm gal}$ is the Galactocentric distance to the cloud. 
\textcolor{black}{$s$ is the size of the cloud ($D\,\tan{(\sqrt{\sigma _l \sigma _b}})$, where $\sigma _l$ and $\sigma _b$ are the intensity-weighted standard deviation values along the Galactic longitude axis and Galactic latitude axis, respectively).}
(This table is available in its entirety in machine-readable form.)
\end{tablenotes}
\end{center}
\end{landscape}

\clearpage
\begin{landscape}
\begin{center}
\begin{table}
  \tbl{Molecular clouds identified in the C$^{18}$O ($J=1-0$) data.}{
  \label{tab:C18O}
  \begin{tabular}{c|cccccccccccccccc}
  \hline              
  ID & $l$ & $b$ & $v_{\rm LSR}$ & $n_{\rm vox}$	 & $\sigma _v$ & $T_{\rm ^{12}CO}$ & $T_{\rm ^{13}CO}$ & $T_{\rm C^{18}O}$ & $w_{\rm ^{12}CO}$ & $w_{\rm ^{13}CO}$ & $w_{\rm C^{18}O}$ & NF & $D$ & $M_{\rm LTE}$ & $R_{\rm gal}$ & $s$\\
  & [$^{\circ}$] & [$^{\circ}$] & [km\,s$^{-1}$] & & [km\,s$^{-1}$] & [K] & [K] & [K] & [K\,km\,s$^{-1}$ & [K\,km\,s$^{-1}$ & [K\,km\,s$^{-1}$ &  & [kpc] & [M$_{\odot}$] & [kpc] & [pc] \\
  &   &   &   & &   &   &   &   &   & degree$^2$] & degree$^2$] & degree$^2$] &  &  &  &  \\
  \hline
  \hline
  C18O-000001 & 10.633 & -0.336 & -3.48 & 1414 & 1.18 & 41.2 & 22.0 & 9.1 & 1.42e-01 & 5.26e-02 & 1.39e-02 & F(1.00) & 16.74 & 383431.88 & 8.86 & 2.98 \\
C18O-000002 & 10.623 & -0.381 & -2.83 & 2630 & 1.51 & 45.3 & 26.5 & 8.4 & 2.85e-01 & 1.02e-01 & 2.31e-02 & F(1.00) & 16.60 & 682457.99 & 8.72 & 3.45 \\
C18O-000003 & 10.210 & -0.320 & 12.12 & 15549 & 2.07 & 13.1 & 11.0 & 7.4 & 7.45e-01 & 3.89e-01 & 1.21e-01 & N(0.12) & 1.87 & 34758.15 & 6.32 & 0.88 \\
  ... &   &   &   & &   &   &   &   &   &   &   &   &  &  &  &  \\
  C18O-006592 & 61.362 & 0.086 & 21.30 & 902 & 0.83 & 45.6 & 20.8 & 3.0 & 1.05e-01 & 3.87e-02 & 4.48e-03 & N(0.17) & 2.05 & 2026.41 & 7.39 & 0.54 \\
C18O-006593 & 60.779 & -0.122 & 20.65 & 681 & 1.13 & 36.7 & 13.8 & 2.2 & 9.92e-02 & 2.99e-02 & 3.15e-03 & N(0.15) & 1.91 & 1330.55 & 7.41 & 0.30 \\
C18O-006594 & 61.674 & 0.287 & 20.65 & 187 & 0.41 & 10.9 & 7.1 & 2.1 & 7.03e-03 & 3.79e-03 & 8.64e-04 & N(0.23) & 2.00 & 155.32 & 7.41 & 0.40 \\
  \hline
  \hline
\end{tabular}}
\end{table}
\begin{tablenotes}
\item[*] $l$, $b$, and $v_{\rm LSR}$ are the peak positions of the C$^{18}$O ($J=1-0$) emission. 
$n_{\rm vox}$ indicates the number of voxels that the cloud contains. 
$\sigma _v$ is the intensity-weighted standard deviation of the velocity. 
$T_{\rm ^{12}CO}$, $T_{\rm ^{13}CO}$, and $T_{\rm C^{18}O}$ are the intensities of the $^{12}$CO ($J=1-0$), $^{13}$CO ($J=1-0$), and C$^{18}$O ($J=1-0$) emission, respectively, at the peak position.
$w_{\rm ^{12}CO}$, $w_{\rm ^{13}CO}$, and $w_{\rm C^{18}O}$ are the combined intensities of the three emissions, respectively.
NF indicates Near (N) or Far (F), which we assigned.
\textcolor{black}{Decimals in parentheses indicate the confidence of the inference (median value of the inference by five models in the all voxels); closer to 0.0 means ``Near'', and closer to 1.0 means ``Far''.}
$D$ is the kinematic distance to the cloud. 
$M_{\rm LTE}$ is the mass of the cloud derived from LTE analysis using the peak brightness temperature of the $^{12}$CO ($J=1-0$) and the brightness temperature of the C$^{18}$O ($J=1-0$). We adopted an abundance ratio [C$^{18}$O]/[H$_2$] of $1.7 \times 10^{-7}$. 
$R_{\rm gal}$ is the Galactocentric distance to the cloud. 
\textcolor{black}{$s$ is the size of the cloud ($D\,\tan{(\sqrt{\sigma _l \sigma _b}})$, where $\sigma _l$ and $\sigma _b$ are the intensity-weighted standard deviation values along the Galactic longitude axis and Galactic latitude axis, respectively).}
(This table is available in its entirety in machine-readable form.)
\end{tablenotes}
\end{center}
\end{landscape}

%\section{Discussion}

\subsection{Mass, size, and velocity dispersion of the clouds}\label{sec:mass}
By assuming local thermodynamic equilibrium (LTE), we determined the column densities of $^{13}$CO for $^{12}$CO clouds and $^{13}$CO clouds.
In addition, we derived the column densities of C$^{18}$O for the C$^{18}$O clouds. 
The equations for the LTE analysis are the same as those used in \citet{1998ApJS..117..387K}, \citet{2014A&A...564A..68S} and \citet{2015ApJS..216...18N}.

\begin{equation}
T_{\rm ex}=\frac{T_0}{\ln[1+(T_0/(T_{\rm peak}(\rm ^{12}CO)+0.84))]}\,[{\rm K}]
\end{equation}

\begin{equation}
J(T)\equiv\frac{1}{\exp(T_0/T)-1}
\end{equation}

\begin{equation}
\tau_{\rm ^{13}CO} = -\ln\left\{1-\frac{T_{\rm MB}(\rm ^{13}CO)}{T_0[J(T_{\rm ex})-0.164]}\right\}
\end{equation}

\begin{equation}
N_{\rm ^{13}CO} = 2.42 \times 10^{14} \left\{\frac{\tau_{\rm ^{13}CO}\Delta V T_{\rm ex}}
{1-\exp[-T_0/T_{\rm ex}]}\right\}
\,[{\rm cm^{-2}}]
\end{equation}

\begin{equation}
\tau_{\rm C^{18}O} = -\ln\left\{1-\frac{T_{\rm MB}(\rm C^{18}O)}{T_0[J(T_{\rm ex})-0.1666]}\right\}
\end{equation}

\begin{equation}
N_{\rm C^{18}O} = 2.42 \times 10^{14} \left\{\frac{\tau_{\rm C^{18}O}\Delta V T_{\rm ex}}
{1-\exp[-T_0/T_{\rm ex}]}\right\}
\,[{\rm cm^{-2}}]
\end{equation}
Here, $T_{\rm ex}$ and $T_{\rm peak}$($^{12}$CO) are the excitation and peak brightness temperatures of the $^{12}$CO voxels in units of K ($T_{\rm MB}$ scale), respectively. 
The same value of $T_{\rm ex}$ was used for each cloud.
$T_{\rm MB}$($^{13}$CO) and $T_{\rm MB}$(C$^{18}$O) are the brightness temperatures in units of K.
$T_0$ are 5.53\,K for Equation (3), 5.29\,K for Equations (5) and (6), and 5.27\,K for Equations (7) and (8), respectively. 
$\Delta V$ is the velocity grid size of the cube, 0.65\,km\,s$^{-1}$. 
The cloud mass was calculated from the total H$_2$ column density.
\begin{equation}
\left(\frac{M_{\rm cloud}}{M_{\odot}}\right) = 4.05\,\times\,10^{-1}\,\mu_{\rm H_2}\,\left(\frac{m_{\rm H}}{\rm kg}\right)\,\left(\frac{d}{\rm pc}\right)^2\,\left(\frac{\Delta l}{\rm arcmin}\right)\,\left(\frac{\Delta b}{\rm arcmin}\right)\,\left(\frac{N_{\rm H_2}}{\rm cm^{-2}}\right), 
\end{equation}
where $\mu_{\rm H_2}$ $\sim \, 2.7$ is the mean molecular weight per H$_2$ molecule, $m_{\rm H}$ is the mass of atomic hydrogen, $d$ is the distance, and $\Delta l$ and $\Delta b$ are the spatial grid sizes.

\begin{figure}
 \begin{center}
  \includegraphics[width=8cm]{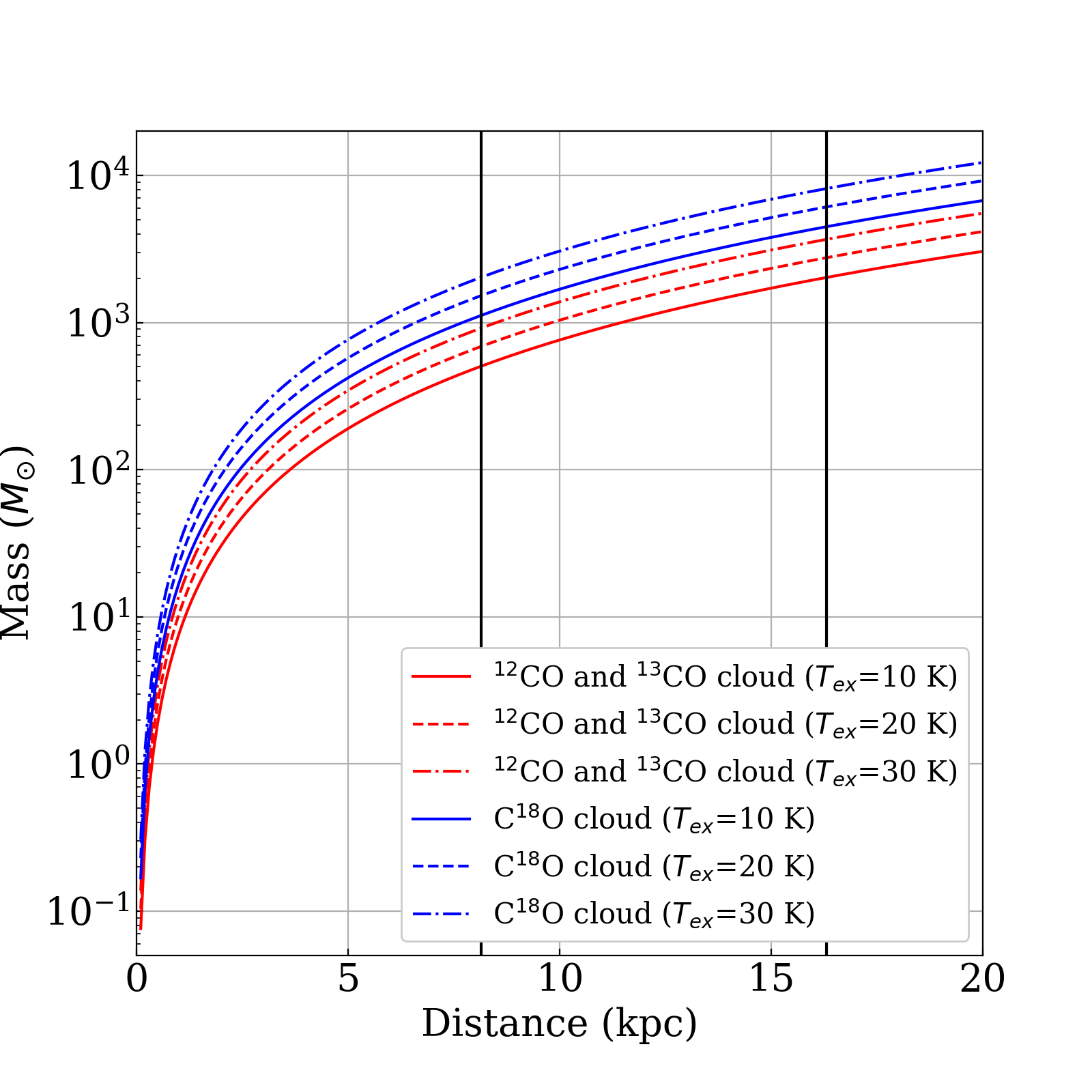} 
 \end{center}
\caption{\textcolor{black}{Detection limit of the cloud mass in this study as a function of distance to clouds. The two thick vertical lines indicate $8.15$\,kpc and $16.3$\,kpc.} }\label{fig:dl}
\end{figure}

\begin{figure}
 \begin{center}
  \includegraphics[width=15cm]{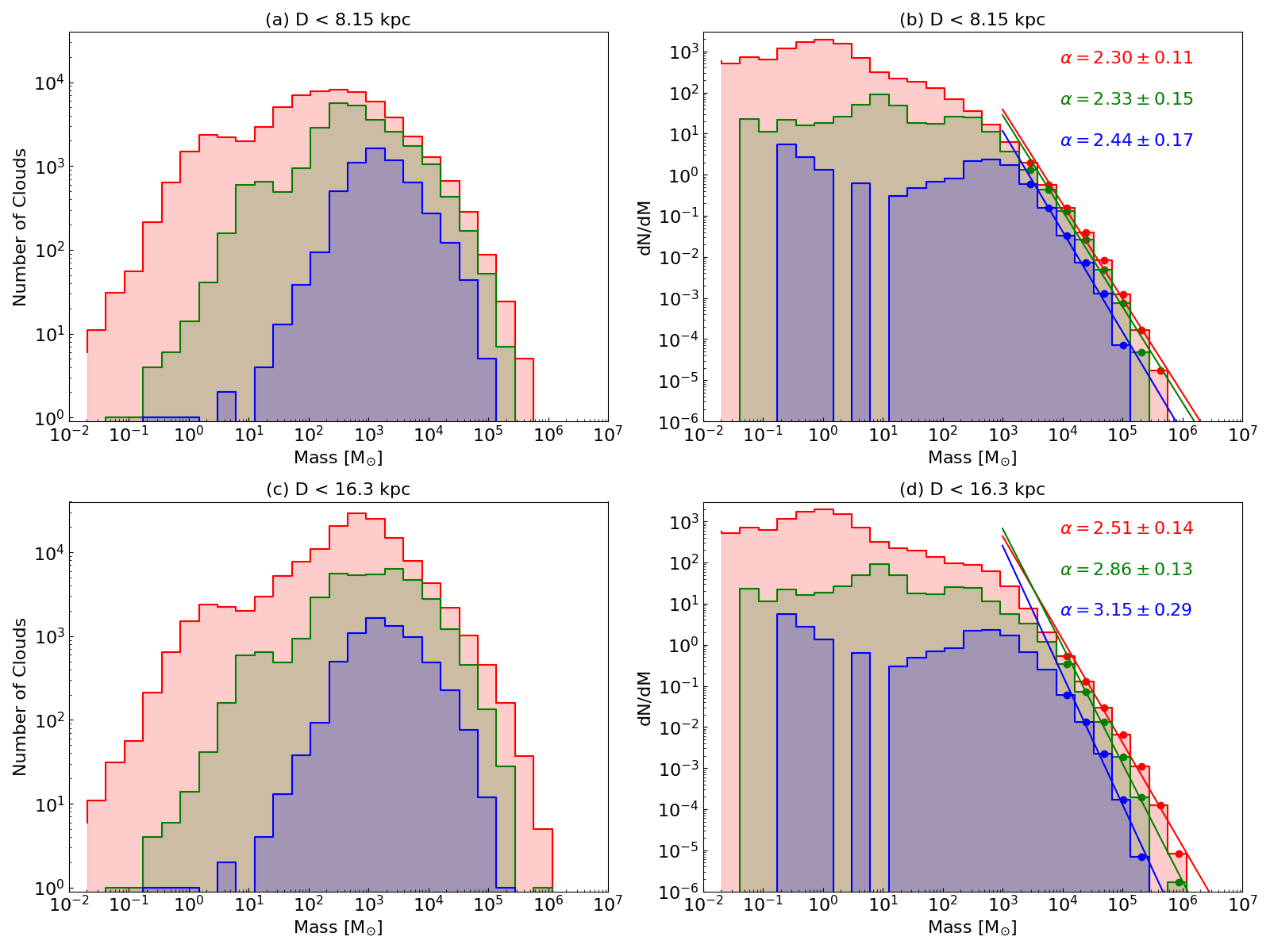} 
 \end{center}
\caption{\textcolor{black}{(a) Histograms of the mass of the $^{12}$CO clouds (red), $^{13}$CO clouds (green), and C$^{18}$O clouds (blue) with a distance of $<$\,8.15\,kpc. (b) Same as the left, but the vertical axis is $dN/dM$. (c) and (d) are the same as (a) and (b), respectively, but with a distance of $<$\,16.3\,kpc.} }\label{fig:hist_mass_all}
\end{figure}

\textcolor{black}{Figure\,\ref{fig:dl} shows the detection limit of the cloud mass in this study as a function of the distance to clouds. Three representative $T_{\rm ex}$ values of 10, 20, and 30 K were adopted.}
\textcolor{black}{Figure\,\ref{fig:hist_mass_all} (a) shows the histograms of the mass of the $^{12}$CO clouds (red), $^{13}$CO clouds (green), and C$^{18}$O clouds (blue) with a distance of $<$\,8.15\,kpc. 
We adopted abundance ratios [$^{13}$CO]/[H$_2$] and [C$^{18}$O]/[H$_2$] of $1.5 \times 10^{-6}$ and $1.7 \times 10^{-7}$, respectively (\cite{1978ApJS...37..407D, 1982ApJ...262..590F}). 
Figure\,\ref{fig:hist_mass_all} (b) is the same as Figure\,\ref{fig:hist_mass_all} (a), but the vertical axis is $dN/dM$.
It appears that $dN/dM \,\propto\,M^{-\alpha}$ in the mass range of $M\,>\,3\times10^3\,M_{\odot}$, whose value is above the mass detection limit shown in Figure\,\ref{fig:dl}.
For the least-squares fitting, $\alpha$ was found to be $2.30\,\pm\,0.11$, $2.33\,\pm\,0.15$, and $2.44\,\pm\,0.17$ for the $^{12}$CO, $^{13}$CO, and C$^{18}$O clouds, respectively. 
Figures\,\ref{fig:hist_mass_all} (c) and (d) are the same as in Figures\,\ref{fig:hist_mass_all} (a) and (b), but with cloud distances of $<$\,$16.3$\,kpc. 
The mass detection limits in $16.3$\,kpc were approximately $1\times10^4\,M_{\odot}$, as shown in Figure\,\ref{fig:dl}. 
Within this mass range, the slopes are steeper than those in Figure \ref{fig:hist_mass_all} (b).}
For example, with respect to the cloud mass function, \citet{2007ApJ...661..830R} reported a similar $\alpha$ value of approximately $2.1$ within an annulus of 2.1\,kpc $<$ $R_{\rm gal}$ $<$ 4.1\,kpc for M33. 
\textcolor{black}{In addition, \citet{2009ApJ...699L.134P} found $\alpha\,=\,2.4$ for the standard CLUMPFIND parameters in the Perseus Molecular Cloud Complex, which has a value that is similar to our results.}

\begin{figure}
 \begin{center}
  \includegraphics[width=12cm]{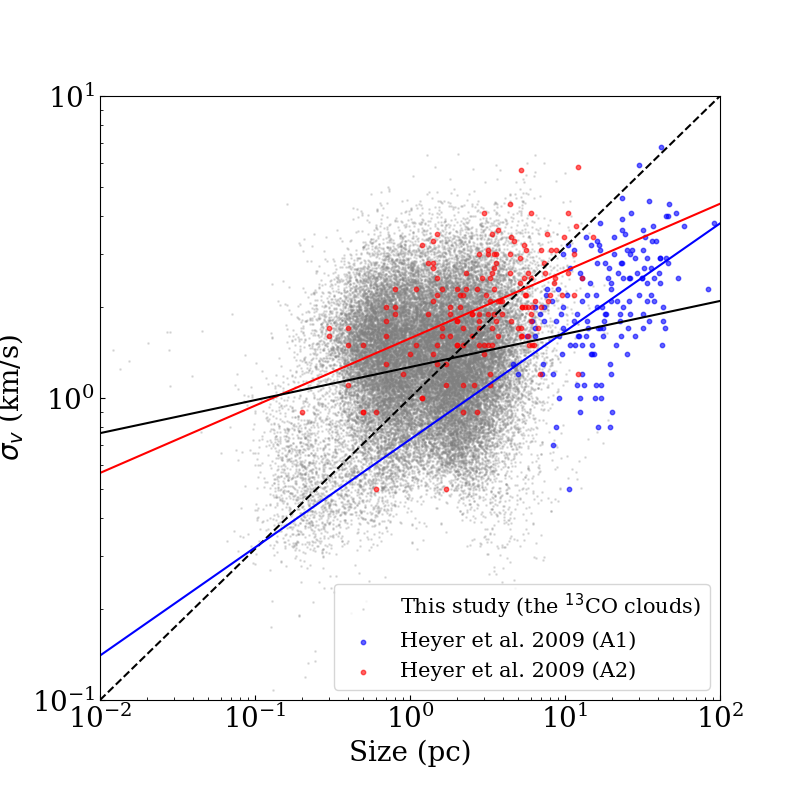} 
 \end{center}
\caption{\textcolor{black}{Relation between the size and $\sigma _v$ of the $^{13}$CO clouds (gray points). The blue and red points show the results of \citet{2009ApJ...699.1092H}. The black, blue, and red solid lines indicate the linear least-squares fit for the gray points, the blue points, and red points in a log-log space, respectively. The dashed-line indicates $\sigma _v = s^{0.5}$, where $s$ is the size of the cloud in units of pc.} }\label{fig:sl}
\end{figure}

\textcolor{black}{Figure\,\ref{fig:sl} shows the scatter plot of the size and $\sigma _v$ of the $^{13}$CO clouds (gray points). 
The definitions of size and $\sigma _v$ are the same as those used in \citet{1987ApJ...319..730S} and \citet{2009ApJ...699.1092H}, and they are the intensity-weighted standard deviation values.
The blue and red points show the results of \citet{2009ApJ...699.1092H} (A1) using the area defined in \citet{1987ApJ...319..730S} and A2, using the area defined by the GRS data). 
\citet{2009ApJ...699.1092H} analyzed the GRS data and re-examined the properties of the Galactic molecular clouds tabulated by \citet{1987ApJ...319..730S}.
The ``Near'' or ``Far'' for most of the clouds in \citet{1987ApJ...319..730S} were determined using a well-matched size-line width relation. 
The black, blue, and red solid lines in Figure\,\ref{fig:sl} indicate the linear least-squares fit for the gray, blue, and red points in a log-log space, respectively.
The slopes were $\sim 0.11$, $\sim 0.36$, and $\sim 0.22$ for the gray, blue, and red points, respectively.
The black dashed line in Figure\,\ref{fig:sl} indicates $\sigma _v = s^{0.5}$, where $s$ is the radius of the cloud in units of PC, indicating the slope of Larson's law (\cite{1981MNRAS.194..809L}). 
There is a size difference between Nobeyama 45-m data (this study) and \citet{2009ApJ...699.1092H}.
This is because of the angular resolution; the difference is small for A2 but significant for A1.
However, $\sigma _v$ increases according to the size, whereas the slope differs among the three.
The slopes tended to become more gradual at higher angular resolutions; However, there may be a tendency for the slope to become steeper as the distance increases.}
In the future, we will analyze our results and discuss them in detail.

%\section{Results}
\section{Face-on-view of the Galaxy}
The distances between the clouds identified above were calculated.
Then, we plotted the face-on-view maps of the column density of $^{13}$CO for the $^{12}$CO clouds and the column density of C$^{18}$O for the C$^{18}$O clouds, which were derived in Section\,\ref{sec:mass}.
Figure\,\ref{fig:faceon_N13CO_in12CO_cf} shows the face-on-view map of the $^{13}$CO column density of the $^{12}$CO clouds and the C$^{18}$O column density of the C$^{18}$O clouds. 
The solar system is the origin, and the cross mark represents the Galactic center.
The tangential points and solar circles are shown by the small and large dotted circles, respectively.
The angular and radial grid sizes of the map are $30''$ and $0.1$\,kpc, respectively.  
In the $^{13}$CO column density map, several spiral arm-like structures and a hole can be observed at ($\theta$, $D$) = ($20^{\circ}$, $7.5$\,kpc). 
This hole is also observed in the map of \citet{2006PASJ...58..847N}.
In contrast, in the C$^{18}$O, from the column density map, the C$^{18}$O cloud is mainly identified as a ring with a radius of $\sim \, 4.5$\,kpc centered on the Galactic center.

\begin{figure}
 \begin{center}
  \includegraphics[width=15cm]{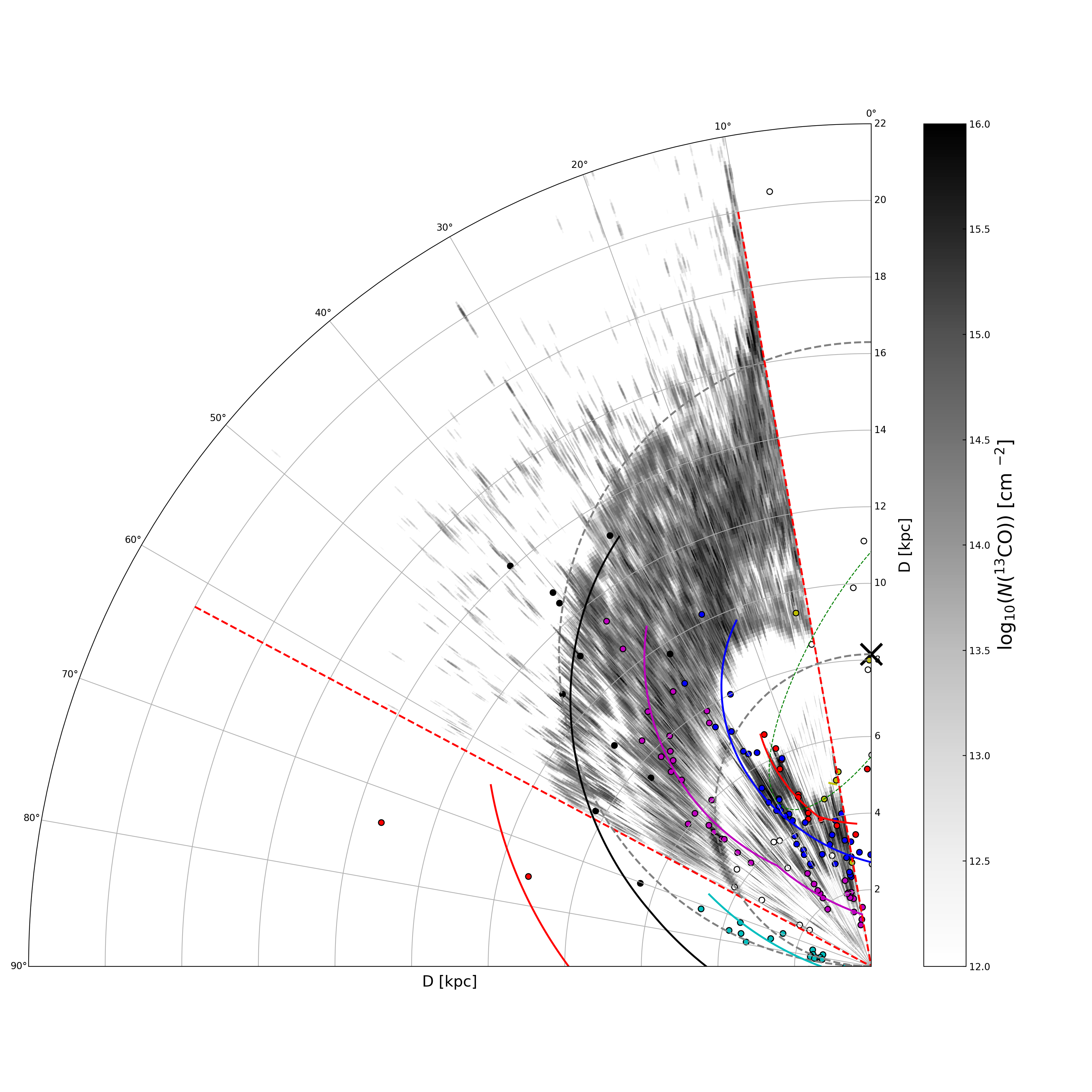} 
 \end{center}
\caption{Face-on-view map of the $^{13}$CO column density of the $^{12}$CO clouds \textcolor{black}{seen from the Galactic North pole}. The circle markers represent the high-mass star-forming regions with measured trigonometric parallaxes (\cite{2019ApJ...885..131R}): 3 kpc arm, yellow; Norma–Outer arm, red; Scutum–Centaurus–OSC arm, blue; Sagittarius–Carina arm, purple; Local arm, cyan; Perseus's arm, black; spurs or sources for which the arm assignment is unclear, white. The ``long'' bar (\cite{2015MNRAS.450.4050W}) is indicated with a green dotted-line ellipse. }\label{fig:faceon_N13CO_Reid_cf}
\end{figure}

Figure\,\ref{fig:faceon_N13CO_Reid_cf} shows a face-on-view map of the $^{13}$CO column density of the $^{12}$CO clouds, and the circles indicate the high-mass star-forming regions with the measured trigonometric parallaxes (\cite{2019ApJ...885..131R}).  
The ``long'' bar (\cite{2015MNRAS.450.4050W}) is shown with a green dotted-line ellipse.
The distribution of molecular gas corresponds well to the majority of high-mass star-forming regions.
Molecular gas is also concentrated in the region of the bar end, which is where the W43 complex is located (\cite{2021PASJ...73S.129K}).

\begin{figure}
 \begin{center}
  \includegraphics[width=15cm]{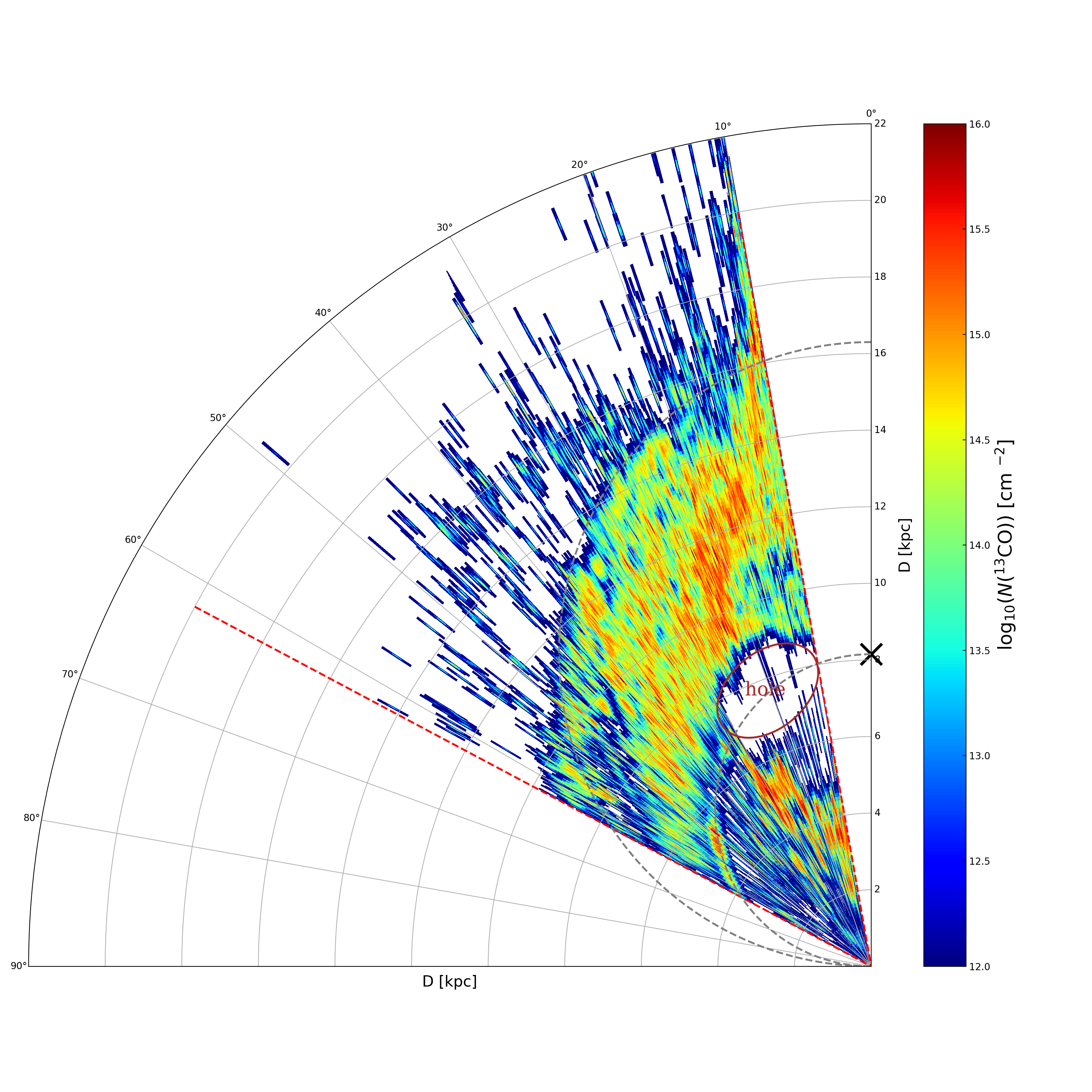} 
 \end{center}
\caption{Face-on-view map of the $^{13}$CO column density of the $^{12}$CO clouds and the C$^{18}$O column density of the C$^{18}$O clouds \textcolor{black}{seen from the Galactic North pole}. 
The solar system is the origin, while the cross mark indicates the Galactic center.
The tangential points and the solar circle are shown by the small and large dotted-line circles, respectively. The Galaxy rotates in a clock-wise direction. The red ellipse shows the hole structure, which is also seen in \citet{2006PASJ...58..847N}}\label{fig:faceon_N13CO_in12CO_cf}
\end{figure}

\begin{figure}
 \begin{center}
  \includegraphics[width=15cm]{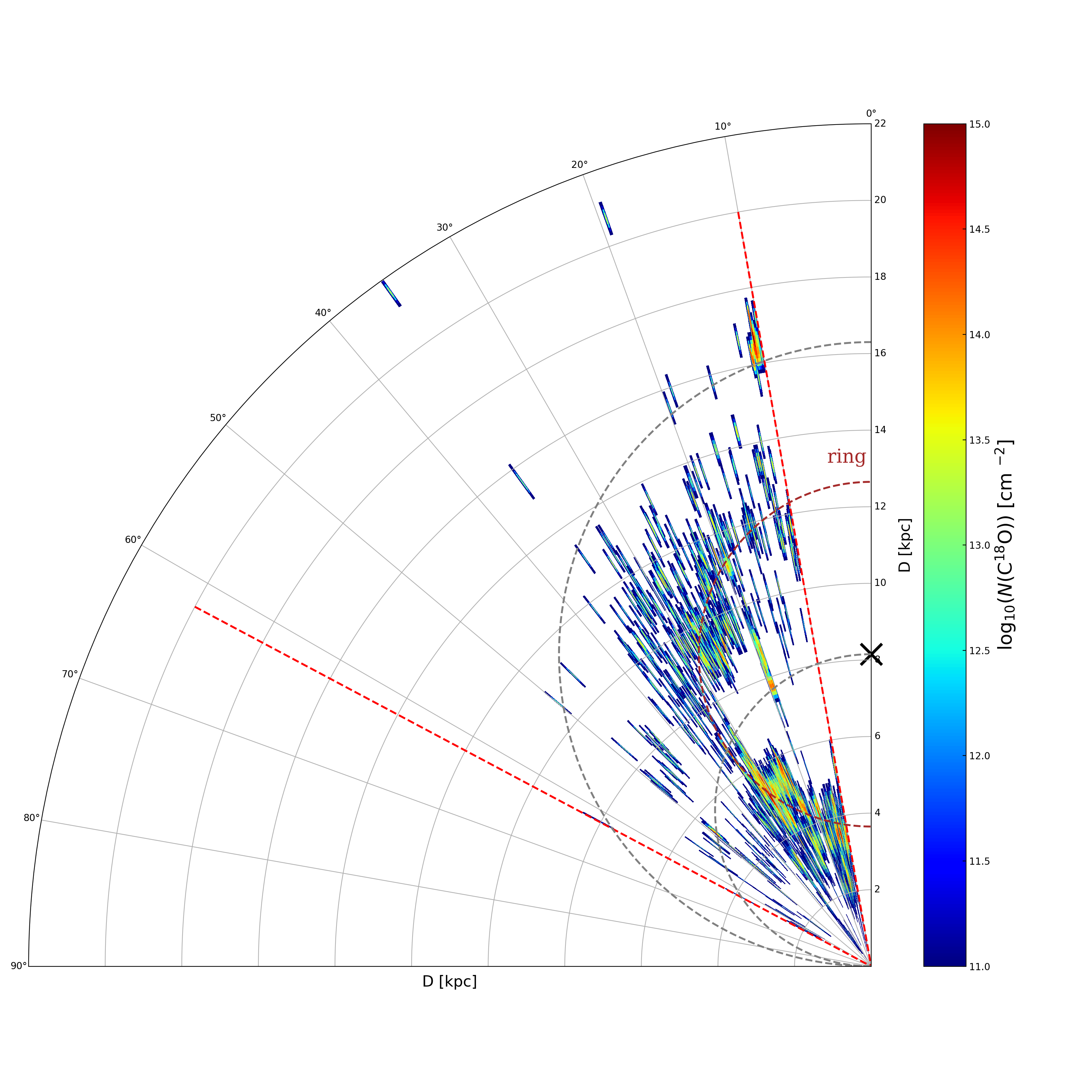} 
 \end{center}
\contcaption{(Continued)}
\end{figure}

\section{\textcolor{black}{Possible errors in our distance estimation}}
\textcolor{black}{In our distance estimation, there are several possible errors that can be attributed to various assumptions and methods.}

\begin{figure}
 \begin{center}
  \includegraphics[width=12cm]{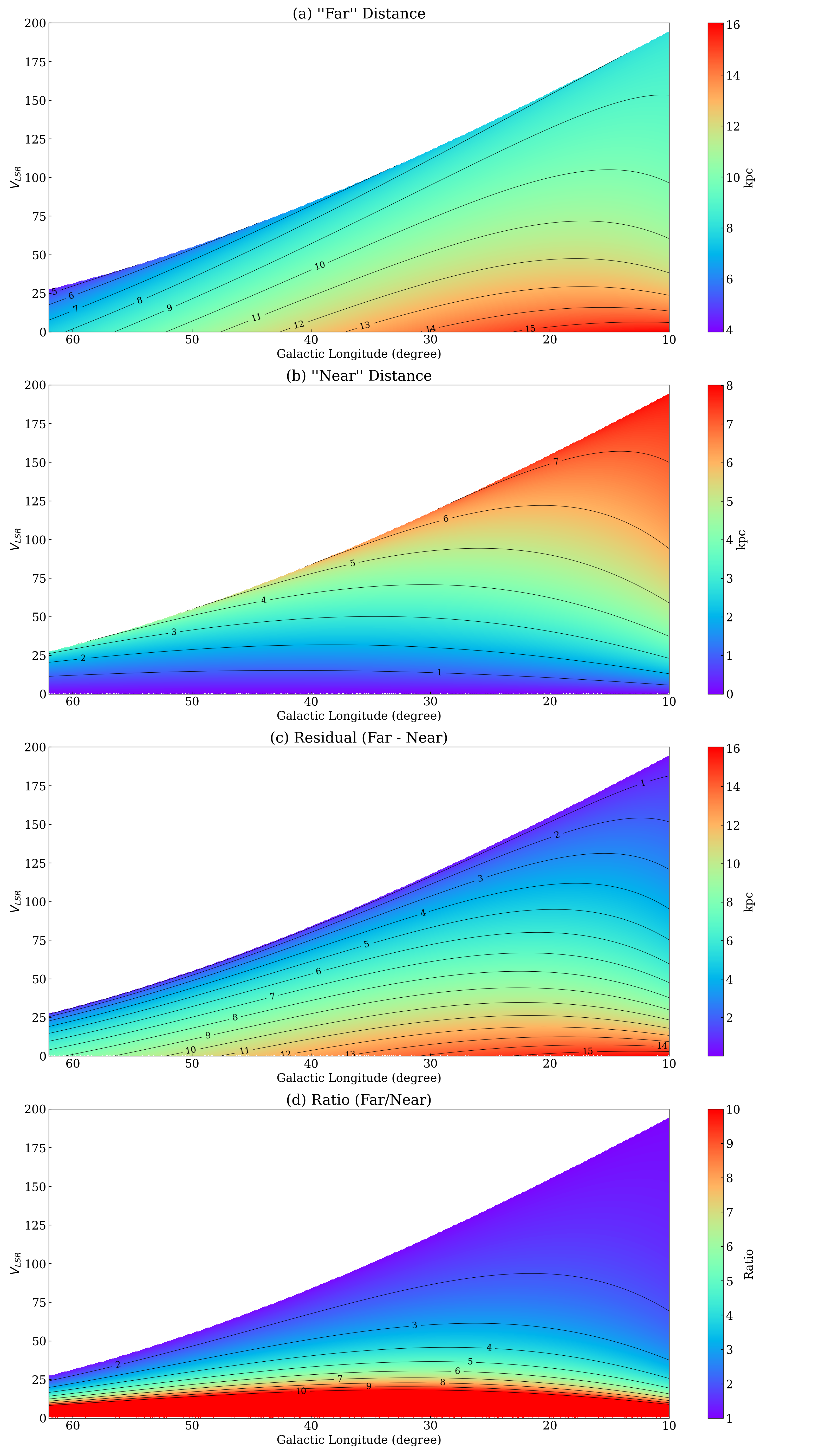} 
 \end{center}
\caption{\textcolor{black}{Kinematic distance in the Galactic plane ($b=0^{\circ}$) for the (a) ``Far'' solution and (b) ``Near'' solution. (c) (d) show their difference and ratio, respectively. }}\label{fig:D_all}
\end{figure}

\textcolor{black}{First, we assumed that all molecular clouds in the Galaxy follow the flat rotation.
However, molecular clouds exhibit intrinsic motions that deviate from the rotation.
The error that results from this is approximately 1 kpc, as discussed in a previous study (\cite{2006PASJ...58..847N}).}
\textcolor{black}{Figures\,\ref{fig:D_all}(a) and (b) show the calculated kinematic distance in the Galactic plane ($b=0^{\circ}$). 
Typically, a deviation of $10-20$\,km\,s$^{-1}$ corresponds to approximately $1$ kpc. 
For example, \citet{1984ApJ...281..624S} reported that the one-dimensional (1D) r.m.s. velocity dispersion of molecular clouds in the Galaxy is $9$\,km\,s$^{-1}$.
Furthermore, there may be a specific velocity depending on the environment in the Galaxy (e.g., the bar end); thus, the hole structure shown in Figure\,\ref{fig:faceon_N13CO_in12CO_cf} may be artificial.
In fact, \citet{2014ApJ...781...89Z} reported that the peculiar motion of W43, which is located near the bar end, is approximately $20$\,km\,s$^{-1}$ toward the Galactic Center by trigonometric parallax measurements of the masers. 
It is very likely that this peculiar motion was induced by the gravitational attraction of the bar.}

\textcolor{black}{Second, the error is due to bias in the teacher (molecular clouds for the dataset with Near--Far annotations) of the model. 
For the teacher dataset, we used only the molecular clouds associated with the H{\sc ii} regions. 
It is possible that they are essentially different from molecular clouds without H{\sc ii} regions, which may affect the results.
However, it is difficult to measure the distances of molecular clouds without H{\sc ii} regions, and we cannot include them in the teacher dataset.}

\textcolor{black}{Third, we assumed that CO emission in a single voxel is derived from a single molecular cloud. 
It is possible that in some regions, ``Near'' clouds and ``Far'' clouds overlap in the Position--Position--Velocity (PPV) space.
However, the angular resolution of the Nobeyama 45-m data that was used is high, and we therefore consider that there is a sufficiently low probability of the ``Near'' and ``Far'' cloud overlapping in the PPV space,} \textcolor{black}{although they cannot be estimated quantitatively. This may be revealed by future detailed numerical simulations of galaxies.}

\textcolor{black}{Fourth, errors in model inference may have affected the results.
The accuracy of the model with respect to the teacher dataset was measured to be 76\%; however, the degree of accuracy for the molecular clouds in the Galactic plane is not currently known.}
\textcolor{black}{Figures\,\ref{fig:D_all}(c) and (d) show the difference and ratio between the ``Far'' and ``Near'' solution. 
The closer to $0$ \\,km \\,s$^{-1}$, the greater is the loss when there were errors in ``Near'' -and ``Far''.
We plan to test the accuracy of the method in this study using pseudo-observational data obtained from numerical simulations of molecular gas in galaxies.}

\textcolor{black}{We checked our decision (``Near'' or ``Far'') for the clouds listed in \citet{1987ApJ...319..730S} (SRBY).
Clouds were selected from only within the region of the data.
Table\ref{tab:cw} lists the number of molecular clouds. 
The number of matches was 125 ($\sim \,66$ \, \%), which is the sum of the cases that both this study and SRBY identify as ``Near'' (98), as well as the case where both of them identify as ``Far'' (27).
In particular, the number of clouds determined to be ``Near'' was higher in our inference than in SRBY.
This result indicates that our inference may be more likely to produce ``Near,” although the angular resolutions of our data and those of SRBY are generally different.
}

\begin{table}
\centering
\begin{tabular}{l|l|c|c|c}
\multicolumn{2}{c}{}&\multicolumn{2}{c}{SRBY}&\\
\cline{3-4}
\multicolumn{2}{c|}{}&Near&Far&\multicolumn{1}{c}{Total}\\
\cline{2-4}
\multirow{2}{*}{This Study}& Near & 98 & 48 & 146\\
\cline{2-4}
& Far & 15 & 27 & 42\\
\cline{2-4}
\multicolumn{1}{c}{} & \multicolumn{1}{c}{Total} & \multicolumn{1}{c}{113} & \multicolumn{    1}{c}{75} & \multicolumn{1}{c}{188}\\
\end{tabular}
\caption{\textcolor{black}{Number of molecular clouds in \citet{1987ApJ...319..730S} (SRBY) distinguished by labels ``Near'' and ``Far'' in this study and SRBY. }}
\label{tab:cw}
\end{table}

\section{Summary}

The main results of this study are summarized as follows:

\begin{enumerate}
\item We presented high angular resolution and wide $^{12}$CO, $^{13}$CO, as well as C$^{18}$O ($J=1-0$) emission data in the 1st quadrant of the Galactic plane obtained with the Nobeyama 45-m radio telescope ($l=62^{\circ}-10^{\circ}$, $|b|< 1^{\circ}$).

\item We attempted to construct a Near--Far inference model using a CNN. In this model, we applied the 3D distribution (position--position--velocity) of the $^{12}$CO ($J=1-0$) emissions as the main input. The training dataset was made from the H{\sc ii} region catalog of the infrared astronomy satellite WISE. Therefore, we were able to construct a CNN model with a 76\% accuracy rate for the training dataset. 
\item Using the CLUMPFIND algorithm, we identified approximately 140,000 clouds in the $^{12}$CO ($J=1-0$) data. By combining this result with the inference of the CNN model, we obtained the distance to the identified clouds. We discovered that the mass of the molecular clouds with a distance of $<\,8.15$\,kpc follows a power-law distribution with an index of approximately $-2.3$ in the mass range $M\,>\,10^3\,M_{\odot}$. Furthermore, the detailed molecular gas distribution of the Galaxy as seen from the Galactic North pole (face-on-view map) was derived. 

\end{enumerate}

\begin{ack}
This study was financially supported by Grants-in-Aid for Scientific Research (KAKENHI) of the Japanese Society for the Promotion of Science (JSPS; grant numbers 17H06740 and JP21H00049) and ``Young interdisciplinary collaboration project'' in the National Institutes of Natural Sciences (NINS). 
The authors would like to thank all members of the FUGIN project. 
Data analysis was performed using Astropy (\cite{2013A&A...558A..33A}) and  APLpy (\cite{2012ascl.soft08017R}). 
The authors would also like to thank NASA for providing the FITS data from the WISE Space Telescope.  
Finally, we wish to express our appreciation to the anonymous reviewer for the insightful comments pertaining to our manuscript.
\end{ack}


\begin{thebibliography}{}
\bibitem[Anderson et al.(2014)]{2014ApJS..212....1A} Anderson, L.~D., Bania, T.~M., Balser, D.~S., et al.\ 2014, \apjs, 212, 1.
\bibitem[Astropy Collaboration et al.(2013)]{2013A&A...558A..33A} Astropy Collaboration, Robitaille, T.~P., Tollerud, E.~J., et al.\ 2013, \aap, 558, A33 
\bibitem[Berry et al.(2007)]{2007ASPC..376..425B} Berry, D. S., Reinhold, K., Jenness, T., Economou, F., 2007, ASP Conf. Ser., 376, 425
\bibitem[Bom et al.(2021)]{2021MNRAS.507.1937B} Bom, C.~R., Cortesi, A., Lucatelli, G., et al.\ 2021, \mnras, 507, 1937. 
\bibitem[Dabeer et al.(2019)]{Sumaiya} Dabeer, S., Khan, M. M., \& Islam, S. 2019, Inform. Med. Unlocked, 16, 100231
\bibitem[Dame et al.(2001)]{2001ApJ...547..792D} Dame, T.~M., Hartmann, D., \& Thaddeus, P.\ 2001, \apj, 547, 792.
\bibitem[Dempsey et al.(2013)]{2013ApJS..209....8D} Dempsey, J.~T., Thomas, H.~S., \& Currie, M.~J.\ 2013, \apjs, 209, 8. 
\bibitem[Dewangan et al.(2020)]{2020MNRAS.496.1278D} Dewangan, L.~K., Baug, T., \& Ojha, D.~K.\ 2020, \mnras, 496, 1278.
\bibitem[Dickman(1978)]{1978ApJS...37..407D} Dickman, R.~L.\ 1978, \apjs, 37, 407
\bibitem[Frerking et al.(1982)]{1982ApJ...262..590F} Frerking, M.~A., Langer, W.~D., \& Wilson, R.~W.\ 1982, \apj, 262, 590 
\bibitem[Fujita et al.(2019)]{2019ApJ...872...49F} Fujita, S., Torii, K., Tachihara, K., et al.\ 2019, \apj, 872, 49. 
\bibitem[Fujita et al.(2021)]{2021PASJ...73S.172F} Fujita, S., Torii, K., Kuno, N., et al.\ 2021, \pasj, 73, S172. 
\bibitem[Fukui et al.(2019)]{2019ApJ...886...14F} Fukui, Y., Tokuda, K., Saigo, K., et al.\ 2019, \apj, 886, 14. 
\bibitem[Heyer et al.(2009)]{2009ApJ...699.1092H} Heyer, M., Krawczyk, C., Duval, J., et al.\ 2009, \apj, 699, 1092.
\bibitem[Jackson et al.(2006)]{2006ApJS..163..145J} Jackson, J.~M., Rathborne, J.~M., Shah, R.~Y., et al.\ 2006, \apjs, 163, 145.
\bibitem[Kawamura et al.(1998)]{1998ApJS..117..387K} Kawamura, A., Onishi, T., Yonekura, Y., et al.\ 1998, \apjs, 117, 387.
\bibitem[Kohno et al.(2018)]{2018PASJ...70S..50K} Kohno, M., Torii, K., Tachihara, K., et al.\ 2018, \pasj, 70, S50.
\bibitem[Kohno et al.(2021)]{2021PASJ...73S.129K} Kohno, M., Tachihara, K., Torii, K., et al.\ 2021, \pasj, 73, S129.
\bibitem[Kohno et al.(2022)]{2022PASJ...74...24K} Kohno, M., Nishimura, A., Fujita, S., et al.\ 2022, \pasj, 74, 24. 
\bibitem[Larson(1981)]{1981MNRAS.194..809L} Larson, R.~B.\ 1981, \mnras, 194, 809. 
\bibitem[Leroy et al.(2021)]{2021ApJS..257...43L} Leroy, A.~K., Schinnerer, E., Hughes, A., et al.\ 2021, \apjs, 257, 43.
\bibitem[Matsuoka et al.(2018)]{Matsuoka} Matsuoka, D., Nakano, M., Sugiyama, D. et al. \ 2018. Prog Earth Planet Sci 5: 80.
\bibitem[M{\`e}ge et al.(2021)]{2021A&A...646A..74M} M{\`e}ge, P., Russeil, D., Zavagno, A., et al.\ 2021, \aap, 646, A74.
\bibitem[Muraoka et al.(2016)]{2016PASJ...68...89M} Muraoka, K., Sorai, K., Kuno, N., et al.\ 2016, \pasj, 68, 89.
\bibitem[Nakanishi \& Sofue(2006)]{2006PASJ...58..847N} Nakanishi, H. \& Sofue, Y.\ 2006, \pasj, 58, 847.
\bibitem[Nishimura et al.(2015)]{2015ApJS..216...18N} Nishimura, A., Tokuda, K., Kimura, K., et al.\ 2015, \apjs, 216, 18.
\bibitem[Nishimura et al.(2018)]{2018PASJ...70S..42N} Nishimura, A., Minamidani, T., Umemoto, T., et al., \ 2018, \pasj, 70, S42.
\bibitem[Pineda et al.(2009)]{2009ApJ...699L.134P} Pineda, J.~E., Rosolowsky, E.~W., \& Goodman, A.~A.\ 2009, \apjl, 699, L134.
\bibitem[Reid et al.(2014)]{2014ApJ...783..130R} Reid, M.~J., Menten, K.~M., Brunthaler, A., et al.\ 2014, \apj, 783, 130.
\bibitem[Reid et al.(2016)]{2016ApJ...823...77R} Reid, M.~J., Dame, T.~M., Menten, K.~M., et al.\ 2016, \apj, 823, 77.
\bibitem[Reid et al.(2019)]{2019ApJ...885..131R} Reid, M.~J., Menten, K.~M., Brunthaler, A., et al.\ 2019, \apj, 885, 131.
\bibitem[Riener et al.(2020)]{2020A&A...640A..72R} Riener, M., Kainulainen, J., Henshaw, J.~D., et al.\ 2020, \aap, 640, A72.
\bibitem[Robitaille \& Bressert(2012)]{2012ascl.soft08017R} Robitaille, T., \& Bressert, E.\ 2012, Astrophysics Source Code Library, ascl:1208.017 
\bibitem[Rosolowsky et al.(2007)]{2007ApJ...661..830R} Rosolowsky, E., Keto, E., Matsushita, S., et al.\ 2007, \apj, 661, 830.
\bibitem[Rosolowsky et al.(2021)]{2021MNRAS.502.1218R} Rosolowsky, E., Hughes, A., Leroy, A.~K., et al.\ 2021, \mnras, 502, 1218.
\bibitem[Schmidhuber(2014)]{2014arXiv1404.7828S} Schmidhuber, J.\ 2014, arXiv:1404.7828
\bibitem[Schuller et al.(2021)]{2021MNRAS.500.3064S} Schuller, F., Urquhart, J.~S., Csengeri, T., et al.\ 2021, \mnras, 500, 3064. 
\bibitem[Shimajiri et al.(2014)]{2014A&A...564A..68S} Shimajiri, Y., Kitamura, Y., Saito, M., et al.\ 2014, \aap, 564, A68. 
\bibitem[Solomon et al.(1987)]{1987ApJ...319..730S} Solomon, P.~M., Rivolo, A.~R., Barrett, J., et al.\ 1987, \apj, 319, 730.
\bibitem[Solomon \& Rivolo(1989)]{1989ApJ...339..919S} Solomon, P.~M. \& Rivolo, A.~R.\ 1989, \apj, 339, 919.
\bibitem[Sorai et al.(2019)]{2019PASJ...71S..14S} Sorai, K., Kuno, N., Muraoka, K., et al.\ 2019, \pasj, 71, S14.
\bibitem[Spitzer(1942)]{1942ApJ....95..329S} Spitzer, L.\ 1942, \apj, 95, 329.
\bibitem[Stark(1984)]{1984ApJ...281..624S} Stark, A.~A.\ 1984, \apj, 281, 624.
\bibitem[Tokuda et al.(2020)]{2020ApJ...896...36T} Tokuda, K., Muraoka, K., Kondo, H., et al.\ 2020, \apj, 896, 36.
\bibitem[Torii et al.(2018)]{2018PASJ...70S..51T} Torii, K., Fujita, S., Matsuo, M., et al.\ 2018, \pasj, 70, S51. 
\bibitem[Torii et al.(2019)]{2019PASJ...71S...2T} Torii, K., Fujita, S., Nishimura, A., et al.\ 2019, \pasj, 71, S2. 
\bibitem[Ueda et al.(2020)]{2020SPIE11452E..2LU} Ueda, S., Fujita, S., Nishimura, A., et al.\ 2020, \procspie, 11452, 114522L.
\bibitem[Umemoto et al.(2017)]{2017PASJ...69...78U} Umemoto, T., Minamidani, T., Kuno, N., et al.\ 2017, \pasj, 69, 78.
\bibitem[Villar et al.(2021)]{2021ApJS..255...24V} Villar, V.~A., Cranmer, M., Berger, E., et al.\ 2021, \apjs, 255, 24.
\bibitem[Wegg et al.(2015)]{2015MNRAS.450.4050W} Wegg, C., Gerhard, O., \& Portail, M.\ 2015, \mnras, 450, 4050.
\bibitem[Williams et al.(1994)]{1994ApJ...428..693W} Williams, J.~P., de Geus, E.~J., \& Blitz, L.\ 1994, \apj, 428, 693.
\bibitem[Zhang et al.(2014)]{2014ApJ...781...89Z} Zhang, B., Moscadelli, L., Sato, M., et al.\ 2014, \apj, 781, 89. 
\end{thebibliography}
\end{document}